\documentclass[prd,aps,tightenlines,superscriptaddress,nofootinbib,showpacs,showkeys,floatfix]{revtex4}
\usepackage{float}
\usepackage{mathrsfs}
\usepackage{amsfonts}
\usepackage{array}
\usepackage{epsfig}
\usepackage{amsmath}    
\usepackage{amssymb}    
\usepackage{graphicx}   
\usepackage{verbatim}   
\usepackage{color}      
\usepackage{subfigure}  
\usepackage{multirow}
\usepackage[letterpaper, portrait, margin=1in]{geometry}
\def\beq{\begin{equation}}
\def\b0{\beta_0}

\def\eeq{\end{equation}}
\def\beeq{\begin{eqnarray}}
\def\eeeq{\end{eqnarray}}

\def\ZcthZ{CT14$_{\rm HERA2}$\ }

\begin{document}
\setlength{\voffset}{0.5in}

\title{CTEQ-TEA parton distribution functions and HERA Run I and II combined data}

\author{Tie-Jiun Hou}
\email{tiejiunh@mail.smu.edu}
\affiliation{
Department of Physics, Southern Methodist University,\\
 Dallas, Texas 75275-0181, USA }
 \author{Sayipjamal Dulat}
\email{sdulat@msu.edu}
\affiliation{
School of Physics Science and Technology, Xinjiang University,\\
 Urumqi, Xinjiang 830046 China }
\affiliation{
Center for Theoretical Physics, Xinjiang University,\\
 Urumqi, Xinjiang 830046 China  }
\affiliation{
Department of Physics and Astronomy, Michigan State University,\\
 East Lansing, Michigan 48824, USA }
 \author{Jun Gao}
\email{jgao@anl.gov}
\affiliation{
Department of Physics and Astronomy,
INPAC, Shanghai Key Laboratory for Particle Physics and Cosmology,\\
Shanghai Jiao-Tong University, Shanghai 200240, China}
\affiliation{
High Energy Physics Division, Argonne National Laboratory,\\
 Argonne, Illinois 60439, USA}
\author{Marco Guzzi}
\email{marco.guzzi@manchester.ac.uk}
\affiliation{
School of Physics and Astronomy, University of Manchester,\\
Manchester M13 9PL, United Kingdom}
\author{Joey Huston}
\email{huston@pa.msu.edu}
\affiliation{
Department of Physics and Astronomy, Michigan State University,\\
 East Lansing, Michigan 48824, USA }
\author{Pavel Nadolsky}
\email{nadolsky@physics.smu.edu}
\affiliation{
Department of Physics, Southern Methodist University,\\
 Dallas, Texas 75275-0181, USA }
\author{Jon Pumplin}
\email{pumplin@pa.msu.edu}
\affiliation{
Department of Physics and Astronomy, Michigan State University,\\
 East Lansing, Michigan 48824, USA }
\author{Carl Schmidt}
\email{schmidt@pa.msu.edu}
\affiliation{
Department of Physics and Astronomy, Michigan State University,\\
 East Lansing, Michigan 48824, USA }
\author{Daniel Stump}
\email{stump@pa.msu.edu}
\affiliation{
Department of Physics and Astronomy, Michigan State University,\\
 East Lansing, Michigan 48824, USA }
\author{ C.--P. Yuan}
\email{yuan@pa.msu.edu}
\affiliation{
Department of Physics and Astronomy, Michigan State University,\\
 East Lansing, Michigan 48824, USA }

\begin{abstract}
  We analyze the impact of the recent HERA Run I+II combination of
inclusive deep inelastic scattering cross-section data
on the CT14 global analysis of parton distribution functions(PDFs).
New PDFs at next-to-leading order and next-to-next-to-leading order,
called CT14$_{\textrm{HERA2}}$, are obtained
by a refit of the CT14 data ensembles,
in which the HERA Run I combined measurements are replaced by
the new HERA Run I+II combination.
The CT14 functional parametrization of PDFs is flexible
enough to allow good descriptions of different flavor combinations,
so we use the same parametrization for CT14$_{\textrm{HERA2}}$
but with an additional shape parameter for describing
the strange quark PDF.
We find that the HERA I+II data can be fit reasonably well, and
both CT14 and CT14$_{\textrm{HERA2}}$ PDFs
can describe equally well the non-HERA data included
in our global analysis.
Because the CT14 and CT14$_{\textrm{HERA2}}$ PDFs agree well within the PDF errors,
we continue to recommend CT14 PDFs for the analysis of LHC Run 2 experiments.
\end{abstract}



\maketitle
\newpage
\tableofcontents
\newpage

\section{Introduction}
\label{sec:Introduction}

CT14 parton distribution functions (PDFs)~\cite{Dulat:2015mca} are
obtained in a global analysis of a variety of hadronic scattering
experimental data. They are suitable for general-purpose QCD calculations
at the Large Hadron Collider (LHC) and in other experiments.
The previous generation of general-purpose PDFs from CTEQ-TEA (CT) group,
designated as
CT10~\cite{Gao:2013xoa,Lai:2010vv},
was used in a wide range of analyses in
hadron collider phenomenology.
The CT10 PDFs were based on diverse experimental data
from fixed-target experiments, HERA and the Tevatron collider,
but without data from the LHC.
The CT14 global analysis
represents the upgrade of the CT10 fit and includes data from the LHC Run I,
as well as updated data from the Tevatron and HERA experiments.
The CT14 PDF sets are available at LHAPDF~\cite{LHAPDF}
together with recent PDF parametrizations from other
groups~\cite{Harland-Lang:2014zoa,Ball:2014uwa,Alekhin:2013nda,Jimenez-Delgado:2014twa}.
The latest version of the PDF4LHC recommendation~\cite{Butterworth:2015oua}
provides users with a consistent procedure on how to combine
the CT14, NNPDF, and MMHT PDF sets in phenomenological analyses.

The CT14 PDFs are determined from data on inclusive high-momentum
transfer processes, for which perturbative QCD is expected to be reliable.
For example, in the case of deep-inelastic lepton scattering (DIS),
only data with $Q>2$ GeV and $W^2>12.5$ GeV$^2$ are used,
where mass squared of the final state hadronic system
$W^2= Q^2({1 \over x} -1)$.
Data in this region are expected to be relatively free of nonperturbative effects,
such as higher-twist or nuclear corrections.
In the global analysis,
the HERA Run I inclusive DIS measurements have imposed
important PDF constraints in the CT10 and CT14 analyses.

In 2015, the H1 and ZEUS collaborations released a novel combination
of measurements of inclusive deep-inelastic
scattering cross sections at $e^{\pm} p$ collider HERA~\cite{Abramowicz:2015mha}.
We refer to this data ensemble as HERA2 throughout this
paper, to be distinguished from the previous combination of HERA data
sets on DIS published in 2009 \cite{Aaron:2009aa}, which we call HERA1.
HERA2 is the combination of HERA Run I measurements of about
100 pb$^{-1}$ of $e^{+}p$ and 15 pb$^{-1}$ of $e^{-}p$ data,
and Run II measurements of 150 pb$^{-1}$ of $e^{+}p$ and 235 pb$^{-1}$ of $e^{-}p$ data,
resulting in a total integrated luminosity of approximately 500 pb$^{-1}$.
The individual H1 and ZEUS measurements used in the combination
were published previously
in Refs.~\cite{Aaron:2009bp,Aaron:2009kv,Adloff:1999ah,Adloff:2000qj,Adloff:2003uh,Aaron:2012qi,Andreev:2013vha,Collaboration:2010ry}
and~\cite{Breitweg:1997hz,Breitweg:2000yn,Breitweg:1998dz,Chekanov:2001qu,Breitweg:1999aa,Chekanov:2002ej,Chekanov:2002zs,Chekanov:2003yv,Chekanov:2003vw,Chekanov:2009gm,Chekanov:2008aa,Abramowicz:2012bx,Collaboration:2010xc,Abramowicz:2014jak}.
The two collaborations employed different experimental techniques
and used different detectors and methods for kinematic reconstruction.
Therefore the new HERA2 {\em combined} measurements exhibit a
significantly reduced systematic uncertainty.

The main goal of this paper is to analyze the impact of
the HERA2 measurements on the CT14 global analysis. We replace the
combined HERA1 data set used in the published CT14
PDFs~\cite{Dulat:2015mca} with the HERA2 set and examine
the resulting changes in PDF central values and uncertainties.
Also, we study the dependence of the goodness of fit
upon kinematic cuts on $Q$ and $x$,
as it was suggested~\cite{Abramowicz:2015mha} that the low-$Q^2$ HERA2 data
are not well fitted by the CT10 and CT14 PDFs. Related studies of the
impact of HERA2 data in the context of MMHT14 and NNPDF3.0 fits
can be found in Refs.~\cite{Harland-Lang:2016yfn,Thorne:2015caa,Rojo:2015nxa}.

To this end, the CTEQ-TEA  PDFs have been refitted at
next-to-leading order (NLO) and next-to-next-to-leading order (NNLO)
by using the global CT14 data ensemble,
but with the HERA2 measurements in place of HERA1.
The new PDFs obtained after the refitting procedure
are named CT14$_\textrm{HERA2}$, to distinguish from CT14.
The HERA2 data set has 1120 data points
in the fitted region with $Q> 2$ GeV and $W^2>12.5$ GeV$^2$.
There are 162 correlated systematic errors,
and seven procedural uncertainties, in addition to the
luminosity uncertainty.
When HERA2 is included in the global fit,
there are in total 3287 data points in the \ZcthZ data ensembles,
compared to 2947 in the original CT14 fits. This is because
two other changes have been made in the data analysis.
 First, we have dropped the New Muon Collaboration (NMC) muon-proton
inclusive DIS data on $F_2^p$~\cite{Arneodo:1996qe},
because that data cannot be fitted well.
As concluded in Ref.~\cite{Pumplin:2002vw},
the NMC $F_2$ proton data are influenced by some unknown
or underestimated systematic errors.
Meanwhile, we continue to include the NMC proton to deuteron ratio data on $F_2^p/F_2^d$.
Second, we updated the data table for the
CMS 7 TeV $5\mbox{ fb}^{-1}$ inclusive jet
experiment~\cite{Chatrchyan:2012bja}, which became available after the
completion of the CT14 study, without appreciable effects on the PDFs.

As in CT14~\cite{Dulat:2015mca},
the theoretical predictions for the majority of processes
in the \ZcthZ fit are calculated at the NNLO level of accuracy.
In particular, a NNLO treatment~\cite{Guzzi:2011ew}
of heavy-quark mass effects in neutral-current(NC) DIS
is realized in the S-ACOT-$\chi$
scheme~\cite{Aivazis:1993kh,Aivazis:1993pi,Collins:1998rz,Tung:2001mv}
and is essential for obtaining correct predictions
for LHC electroweak cross sections~\cite{Gao:2013wwa,Lai:2010nw,Nadolsky:2008zw,Tung:2006tb}.
However, the calculations for charged-current(CC) DIS
and inclusive jet production are included at NLO only;
in both cases, the complete NNLO contributions are not yet
available. In Sec. II of Ref.~\cite{Dulat:2015mca}, we presented
various arguments suggesting that the expected impact of
the missing NNLO effects in jet production on the PDFs
is small relative to current experimental errors. Similarly, the
NNLO contribution to charged-current DIS, including massive charm
scattering contributions \cite{Berger:2016inr},
is modest compared to the experimental uncertainties.

It is useful to review quickly the advances in the CT14 global analysis,
compared to CT10.
{\em Regarding data:}
The new LHC measurements of $W^{\pm}$ and $Z^{0}$
cross sections~\cite{Aad:2011dm,Chatrchyan:2013mza,Aaij:2012vn}
directly probe flavor separation
of $u, \overline{u}$ and $d, \overline{d}$ partons
in an $x$-range around $0.01$ that was not directly assessed
by earlier experiments.
The updated measurements of electron charge asymmetry
from the D\O~ collaboration~\cite{D0:2014kma}
probe the $d$ quark PDF at $x>0.1$.
These measurements are included in the CT14 and \ZcthZ analyses.
{\em Regarding parametrization:}
In the CT14 analysis, the description of variations
in relevant PDF combinations, such as $d(x,Q)/u(x,Q)$
and $\bar d(x,Q)/\bar u(x,Q)$, is improved, as compared to CT10,
by increasing the number of free PDF parameters from 25 to 28.
The functional form for the initial scale PDFs adopted by the
CT14 fit is parametrized by Bernstein polynomials
(reviewed in the Appendix of Ref.~\cite{Dulat:2015mca})
which have the property that a single polynomial is dominant
in any given $x$ range,
hence reducing undesirable correlations among the PDF parameters
that sometimes occurred in CT10.
Also, in the asymptotic limits of $x \rightarrow 0$ or $x\rightarrow 1$,
the CT14 functional forms allow the ratios of
$d/u$ or $\bar d/\bar u$ to reach any values,
so that these ratios are determined by the global fit;
this is in contrast to the more constrained behavior
of those PDF ratios assumed in the CT10 parametrization forms.

The \ZcthZ fit adopts the same functional form for the initial scale
parametrization as CT14, except for the strange quark and antiquark PDFs.
More specifically, in the
CT14$_\textrm{HERA2}$ analysis,
we have used the CT14 PDF functional form~\cite{Dulat:2015mca}
at the initial scale $Q_0 = 1.3\, {\rm GeV}$,
\begin{equation}\label{eq:par}
 x \, f_a(x,Q_0) \, = \, x^{a_1} \, (1-x)^{a_2} \,P_a(x),
\end{equation}
where the $P_a(x)$ functions are linear combinations of
Bernstein polynomials.
In the CT14 fit\ \cite{Dulat:2015mca}, the strange quark PDF
is parametrized according to Eq.~(\ref{eq:par}),
with $P_{s}(x)$ being a constant.
There, we have tied $a_{1}$ to the common $a_{1}$ of $\bar{u}$ and $\bar{d}$,
and assumed $s(x) = \bar{s}(x)$ in the analysis.
Thus, we have just two parameters for the strange quark and antiquark PDFs
in our standard CT14 analysis: $a_{2}$ and normalization.
With this limitation on $s(x,Q_{0})$, we find that it is necessary
to extend the strange quark uncertainty by
adding two ``extreme strange'' PDFs to the set of Hessian error PDFs.
In the CT14$_\textrm{HERA2}$ PDFs, we use a different technique
to avoid underestimating the strangeness uncertainty
provided by the Hessian error PDF set: while in the published CT14
PDFs, we set $a_1(s)=a_1(\bar s)= a_1(\bar d) = a_1(\bar u)$;
in the \ZcthZ fit, we allow $a_1(s)=a_1(\bar s)$ to differ from
$a_1(\bar d) = a_1(\bar u)$.
By freeing the parameter $a_{1}(s)$,
we find that it is not necessary to construct additional extreme
strange quark PDFs.
So, whereas the CT14 error PDFs include two extreme strange and
two extreme gluon PDFs,
the CT14$_\textrm{HERA2}$ error PDFs include
only two extreme gluon PDFs to model the
uncertainty of gluon PDFs in the very small $x$ region.
Thus the total number of error PDFs is the same
for CT14 and CT14$_\textrm{HERA2}$, {\it viz.} 56 error PDFs.

To summarize, we use this parametrization,
differing from the standard CT14 parametrization \cite{Dulat:2015mca}
only by the addition of one free parameter for $s(x,Q_{0})$;
and we refit the CT14 data set,
with the HERA1 combined data replaced by the HERA2 combination,
after dropping the NMC muon-proton inclusive DIS data on
$F_{2}^{p}$~\cite{Arneodo:1996qe}
 and correcting the data table for the
CMS 7 TeV $5\mbox{ fb}^{-1}$ inclusive jet experiment~\cite{Chatrchyan:2012bja}.

The rest of the paper summarizes findings of
the \ZcthZ global analysis, presented in several parts:

\begin{itemize}
\item{Section 2 concerns the goodness-of-fit
for this new QCD global analysis
with special emphasis on the quality of the fit
to the HERA2 combined data.
We find a large value of $\chi^{2}/N_{pts}$ for a subset of the
HERA2 measurements, from $e^{-}p$ scattering,
and we discuss the origin of this increase.}
\item{Section 3 describes a study of the role of
HERA2 data points at low $Q$.
This is studied by excluding low-$Q$ data points
and refitting the PDFs.}
\item{Section 4 concerns the changes of the PDFs themselves.
We find some changes from CT14 to \ZcthZ, but they are not
significant within the standard
CTEQ estimates of PDF uncertainties.}
\item{Section 5 is a summary of our conclusions.}
\end{itemize}

In the end, we find that the differences between
CT14$_\textrm{HERA2}$ and CT14 PDFs are smaller than
the uncertainties of the PDFs,
as estimated by the Hessian method of error propagation.
For this reason we reckon that the standard CT14 PDFs should continue
to be used for making predictions to compare against current
and future LHC data. However, we will make the
CT14$_\textrm{HERA2}$ PDFs available
in the LHAPDF format for specialized studies, such as those that are
sensitive to behavior of strange (anti)quark PDFs.
\section{The Global analysis with the final HERA2 combined Data
\label{sec2}}

As we explained in the introduction, when constructing a PDF ensemble
for general-purpose applications, the CTEQ-TEA global analysis
selects the experimental data points at large enough $Q$ and $W$, where
contributions beyond the leading-twist QCD are reduced.
With the default lower $Q$ cut on the selected data points,
$Q \geq Q_{\textrm{cut}} = 2$ GeV, the HERA1 ensemble contains 579
data points, while that of HERA2 contains 1120 data points.
In Table \ref{tbl:tbl1} we summarize the results for the total $\chi^2$ values
of the HERA1 combined data (column 2) and HERA2 combined data (column 3),
for both NLO and NNLO approximations of QCD.
The rows CT14(NLO) and CT14(NNLO) use the published CT14 PDFs,
with no refitting; they were fit with HERA1 data.
The rows NLO10, NLO55, NNLO10 and NNLO55 are refits with
 a slightly more flexible parametrization for the
 strange quark PDF and the inclusion of the
non-HERA data sets,
as described in Sec. I;
NLO10 and NNLO10 use only HERA1 data;
NLO55 and NNLO55 use HERA1 data with weight 0.5 and
HERA2 data with weight 0.5 in the global $\chi^{2}$ sum.
The rows ${\rm CT14}_{\rm HERA2(NLO)}$ and ${\rm CT14}_{\rm HERA2(NNLO)}$
use the same parametrization and non-HERA data as NLO10 and NNLO10,
but they use only HERA2 data.
Note that $\chi^{2}_{\rm HERA1}$ increases, and $\chi^{2}_{\rm HERA2}$ decreases,
as we vary the balance of HERA1 and HERA2 data used in the analysis,
from weights $\{1,0\}$ to $\{0.5,0.5\}$  to $\{0,1\}$.
However, the changes are not large, given the number of data points,
579 and 1120 respectively.
We have also compared the $\chi^{2}$ values for {\em non-HERA data}
for the new fits, and we find that $\chi^{2}_{\rm non-HERA}$ is
essentially unchanged as we vary the balance of HERA1 and HERA2 data,
with the three weighting choices.
This shows that
the HERA1 and HERA2 data sets are equally consistent with the non-HERA data.

\begin{table}[h]
\begin{tabular}{|l|c|c|}
\hline
    & $\chi^2_{\textrm{HERA1}}$ (wt); $N_{\textrm{pts}}=579$     & $\chi^2_{\textrm{HERA2}}$ (wt); $N_{\textrm{pts}}=1120$
\tabularnewline
\hline
\hline
CT14(NLO)                   &  590       &  1398        \tabularnewline 
\hline
NLO10                       &  576 (1.0) &  1404 (0.0)  \tabularnewline 
NLO55                       &  586 (0.5) &  1374 (0.5)  \tabularnewline 
CT14$_\textrm{HERA2(NLO)}$  &  595 (0.0) &  1373 (1.0)   \tabularnewline 
\hline
\hline
CT14(NNLO)                  &  591       &  1469        \tabularnewline 
\hline
NNLO10                      &  583 (1.0) &  1458 (0.0)  \tabularnewline 
NNLO55                      &  596 (0.5) &  1411 (0.5)  \tabularnewline 
CT14$_\textrm{HERA2(NNLO)}$ &  610 (0.0) &  1402 (1.0)   \tabularnewline 
\hline
\hline
\end{tabular}
\caption{$\chi^2$ values for the HERA Run I data set ($\equiv$ HERA1)
and the HERA Run I+II combined data set ($\equiv$ HERA2).
The CT14 NLO and NNLO results use the published CT14 PDFs, i.e., without refitting.
The other results are fits made with weights $\{1,0\}$, $\{0.5,0.5\}$ or $\{0,1\}$
for the HERA1 and HERA2 data sets, respectively.
[The $\{1,0\}$ fits are not identical to CT14 because they were made
(i) with a slightly more flexible parametrization for the strange quark PDF,
(ii) without the NMC $F_{2}^{p}$ measurements,
and (iii) with an updated data table for CMS jet production.]
\label{tbl:tbl1}}
\end{table}

Furthermore, we find that
the NLO fit has a lower value of global $\chi^2$ than the NNLO fit.
This is a robust result: it is independent of
whether a HERA1 or HERA2 data set is used.
It is also still true if $\alpha_s(m_Z)$, $m_b$, and $m_c$ are varied as free
parameters---separately, of course, for NLO and NNLO.
The conclusions still hold if the kinematic cut $Q_{\textrm{cut}}$ is
raised, cf. Sec.~\ref{kinematic-cuts}.

\begin{figure}[htbp]
\includegraphics[width=0.40\textwidth]
{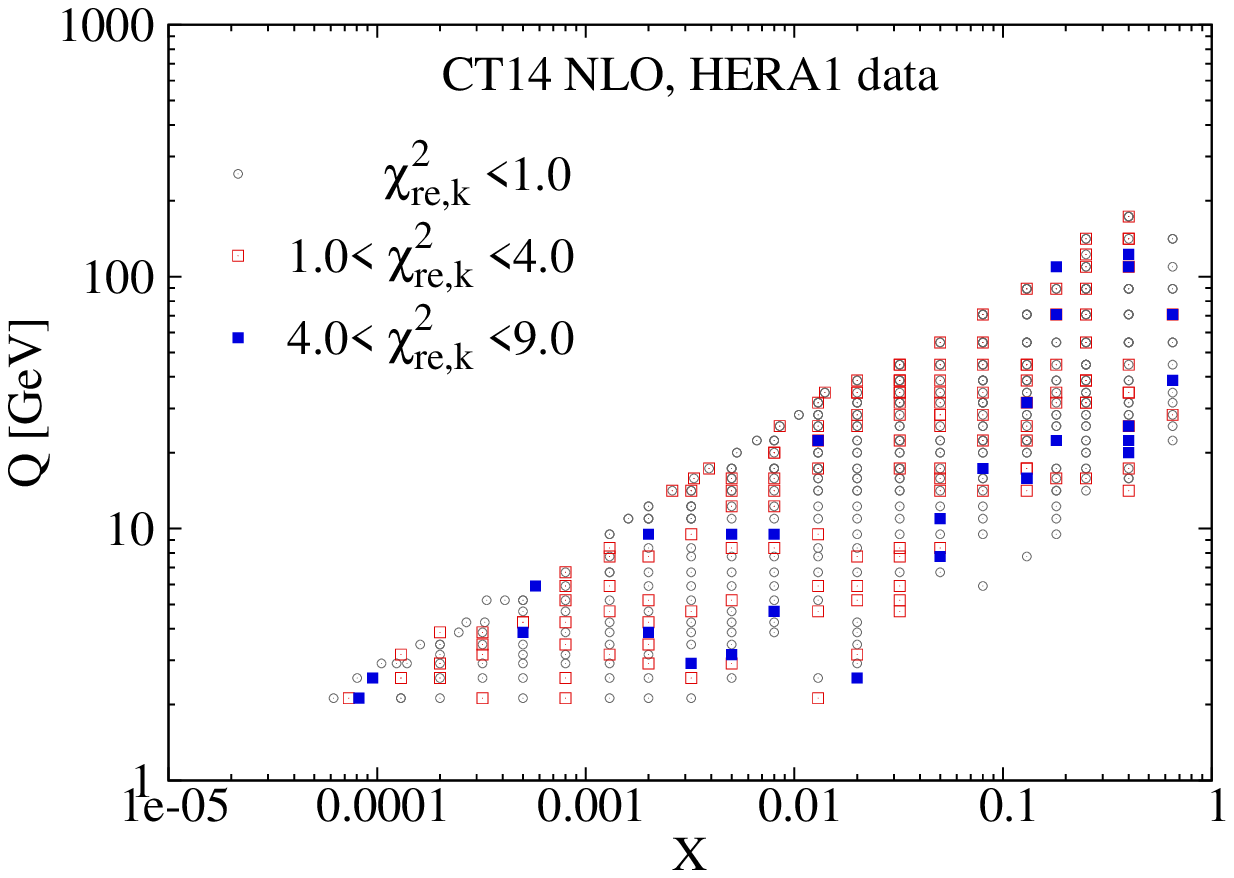}
\includegraphics[width=0.40\textwidth]
{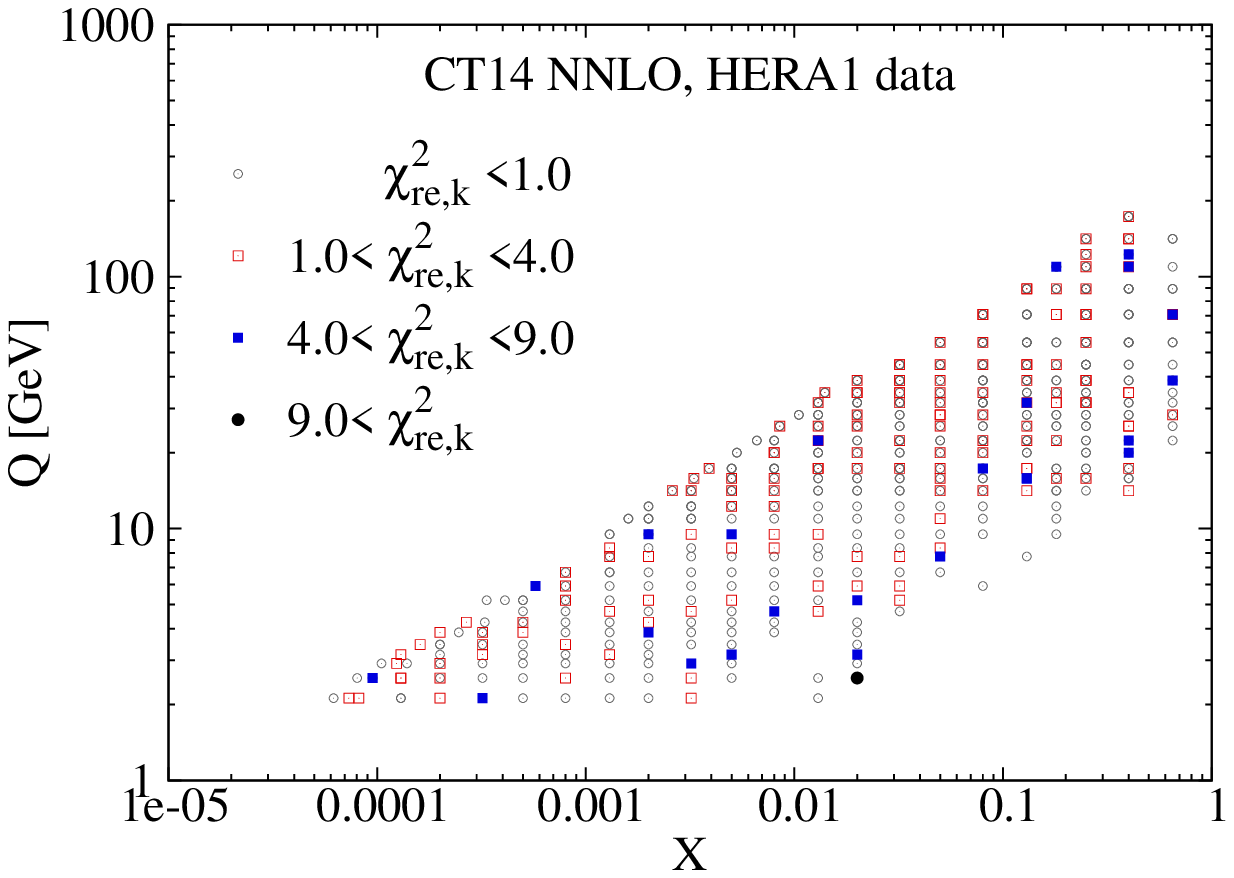}
\includegraphics[width=0.40\textwidth]
{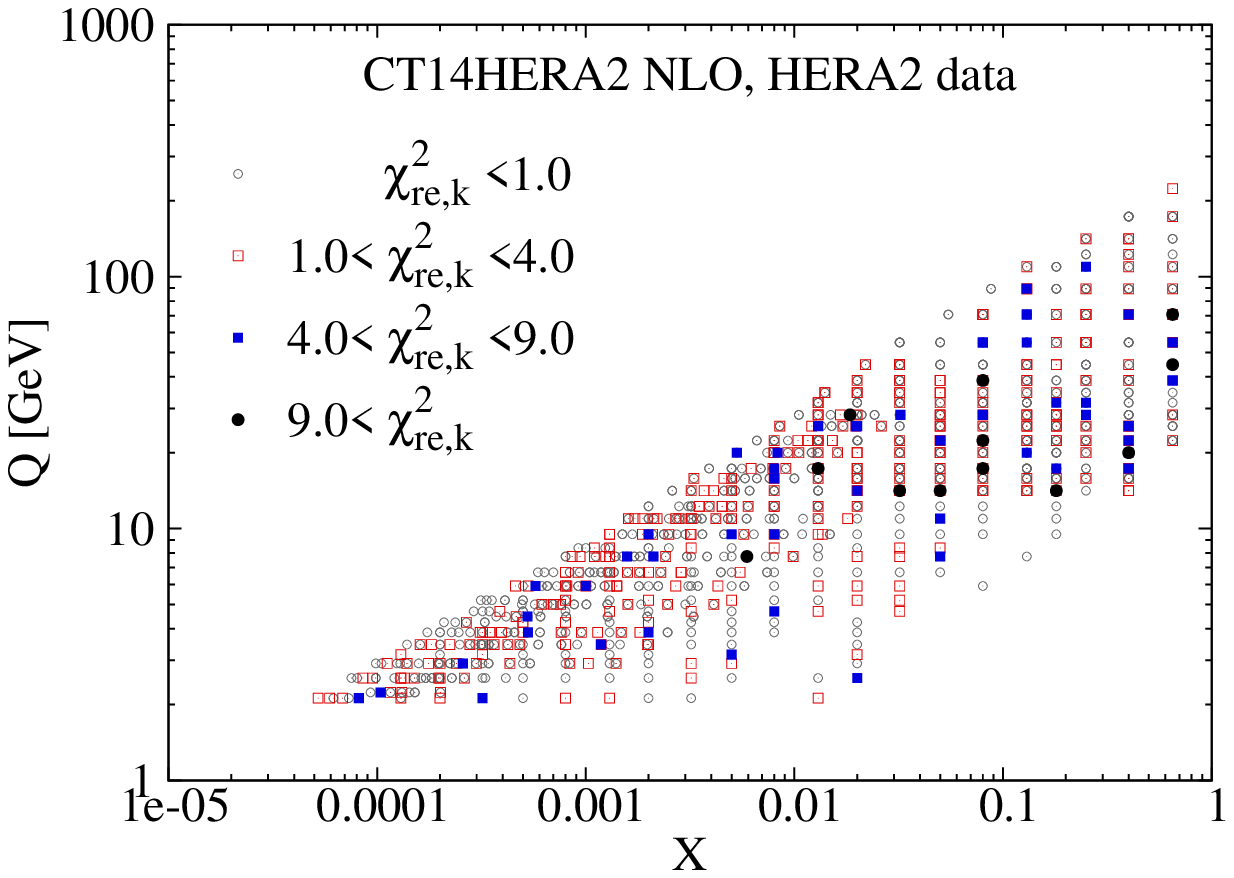}
\includegraphics[width=0.40\textwidth]
{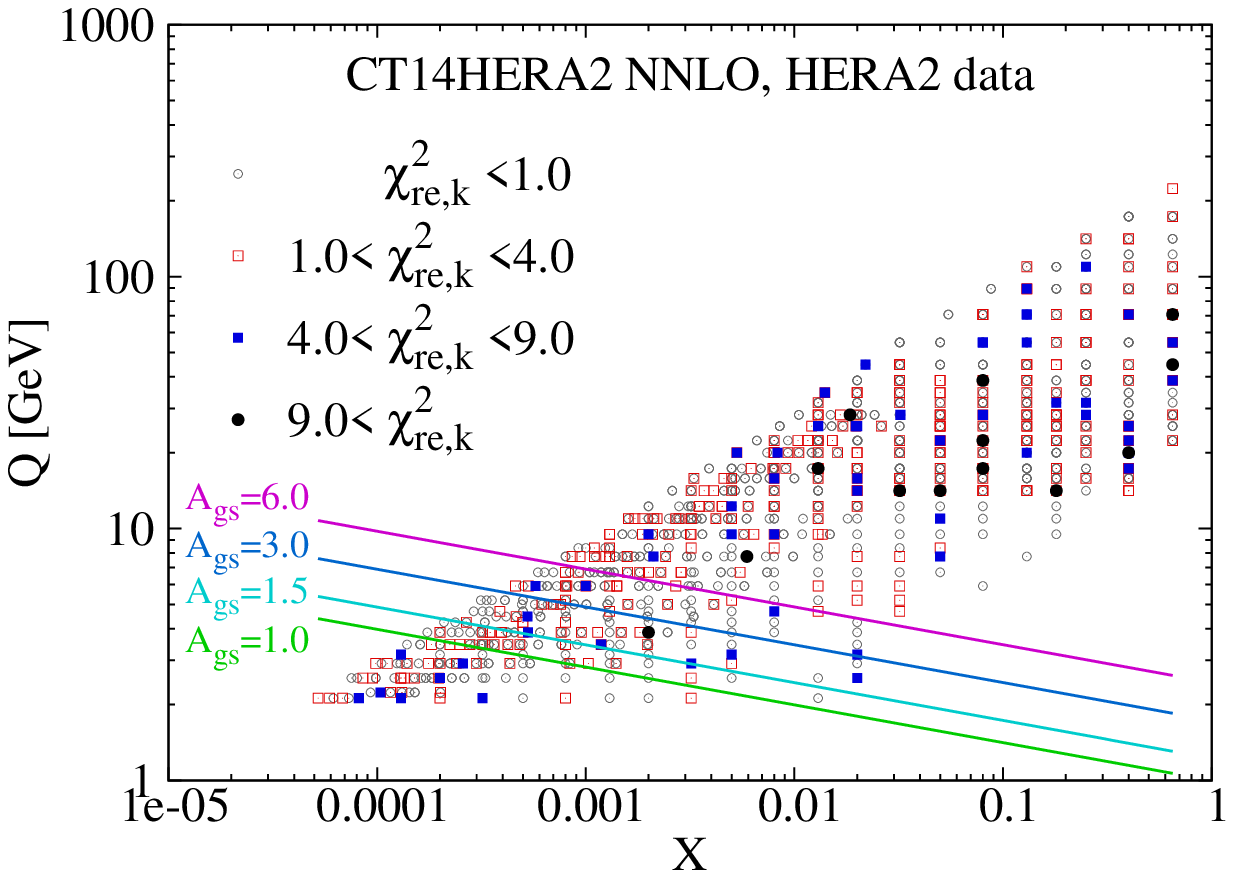}
\caption{The distribution of $\chi^2_{{\rm re},k}$ of HERA1 and
HERA2 ensembles in the $(x,Q)$ plane,
for the CT14 (upper row) and CT14$_\textrm{HERA2}$ (lower row) fits, respectively.}
\label{chi2res}
\end{figure}

In order to understand the impact of the HERA2 data,
we focus on some more detailed quantitative studies in Figs. 1-3.
Considering the value of the {\em global} $\chi^2$ per number of points ($N_{pts}$),
i.e., the {\em overall} goodness of fit for the QCD global analysis,
we find $\chi^2/N_{pts}$ to be 1.07 and 1.09, respectively, at the NLO and NNLO,
which is about the same as for the standard CT14 global analysis~\cite{Dulat:2015mca}.
However,
the values of $\chi^2_{\rm HERA2}/N_{pts}$ for the HERA2 data after refitting
are found to be 1.22 and 1.25, respectively, at the NLO and NNLO.
(For comparison, the $\chi^2_{\rm HERA1}/N_{pts}$ for the HERA Run I ensemble data in the CT14 fits
is about 1.02 at either NLO or NNLO.)
These large values of $\chi^{2}_{\rm HERA2}/N_{pts}$
raise a question: do they come from a few isolated data points, or
from a systematic difference between data and theory?
To address this question, in Fig.~\ref{chi2res}
we show the distribution of the reduced-$\chi^{2}$
($\equiv \chi^{2}_{\rm re}$) values for individual data points,
as they are distributed over the $(x,Q)$ kinematic plane.

The definition of $\chi^{2}_{\rm re}$ is,
for an individual data point ($k$),
\begin{equation}\label{eq:reX2}
\chi^{2}_{{\rm re},k} =
(D_{k} - T_{k} - \sum_{\alpha} \lambda_{\alpha}\beta_{k\alpha})^2/s_{k}^{2},
\end{equation}
where $D_{k}$ is the central data value,
$T_{k}$ is the theory value,
$s_{k}$ is the uncorrelated error,
and the sum over $\alpha$ is an effective shift in the central value
$D_k$ caused by optimized systematic nuisance parameters
$\lambda_\alpha$.
[See, e.g.,  Eq.\ (4) in the original CT10 analysis.~\cite{Lai:2010vv}.]
Thus, $\chi^{2}_{{\rm re}, k}$
represents our best measure for the difference
between data and theory for the $k$th data point.
The total $\chi^2_{exp}$ for the experimental data set quoted in
Table~\ref{tbl:tbl1} (where exp stands for HERA1 or HERA2)
is obtained by summing $\chi^{2}_{{\rm re}, k}$
over all experimental points and adding the penalty $R^2$ for
deviations of the optimized nuisance parameters $\lambda_\alpha$
from their central values at 0,
\begin{equation}
  \chi^2_{exp} = \sum_{k=1}^{N_{pts}} \chi^{2}_{{\rm re}, k} +
  \sum_{\alpha}\lambda^2_{\alpha} \equiv  \chi^{2}_{{\rm re}} + R^2.
  \label{chi2}
\end{equation}

To identify the source of the elevated total $\chi^{2}$ for the HERA2
ensemble,
we first scrutinize the contributions $\chi^{2}_{{\rm re}, k}$ from
the individual points. Figure~\ref{chi2res}
illustrates the values of $\chi^{2}_{{\rm re},k}$
when the HERA1 data are compared to the CT14 (NLO and NNLO) theory,
and the HERA2 data are compared to CT14$_\textrm{HERA2}$ (NLO and NNLO) theory.
The bottom-right inset also shows different values of the
geometric scaling variable $A_{gs}$ that are discussed
in Sec.~\ref{kinematic-cuts}.
In the subfigures for HERA2 (either at NLO or NNLO),
we notice that points with $\chi^{2}_{{\rm re},k} > 4$
are rather uniformly distributed throughout the
$(x,Q)$ phase space, without being concentrated in a
particular region.
In other words, the elevated values of $\chi^2_{\rm HERA2}$
in Table~\ref{chi2res} do not
arise from a single $(x,Q)$ kinematic region.

\setcounter{subsubsection}{0}
\subsubsection{Varied statistical weights for the HERA2 data}

An interesting way to assess the impact of the HERA2 combined data
is to {\em vary the weight} given to this data set in the global
$\chi^{2}$ function.
Namely, we increase the
statistical weight $w$ of the HERA2 data; that is, we include
$w\cdot \chi^2_{\rm HERA2}$, with $w > 1$,
instead of the default $\chi^2_{\rm HERA2}$ (with $w = 1$),
into the global function $\chi^2$.
The purpose here is to see whether increasing the HERA2 weight
will induce large changes in the PDFs.

\begin{figure}
\includegraphics[width=0.40\textwidth]
{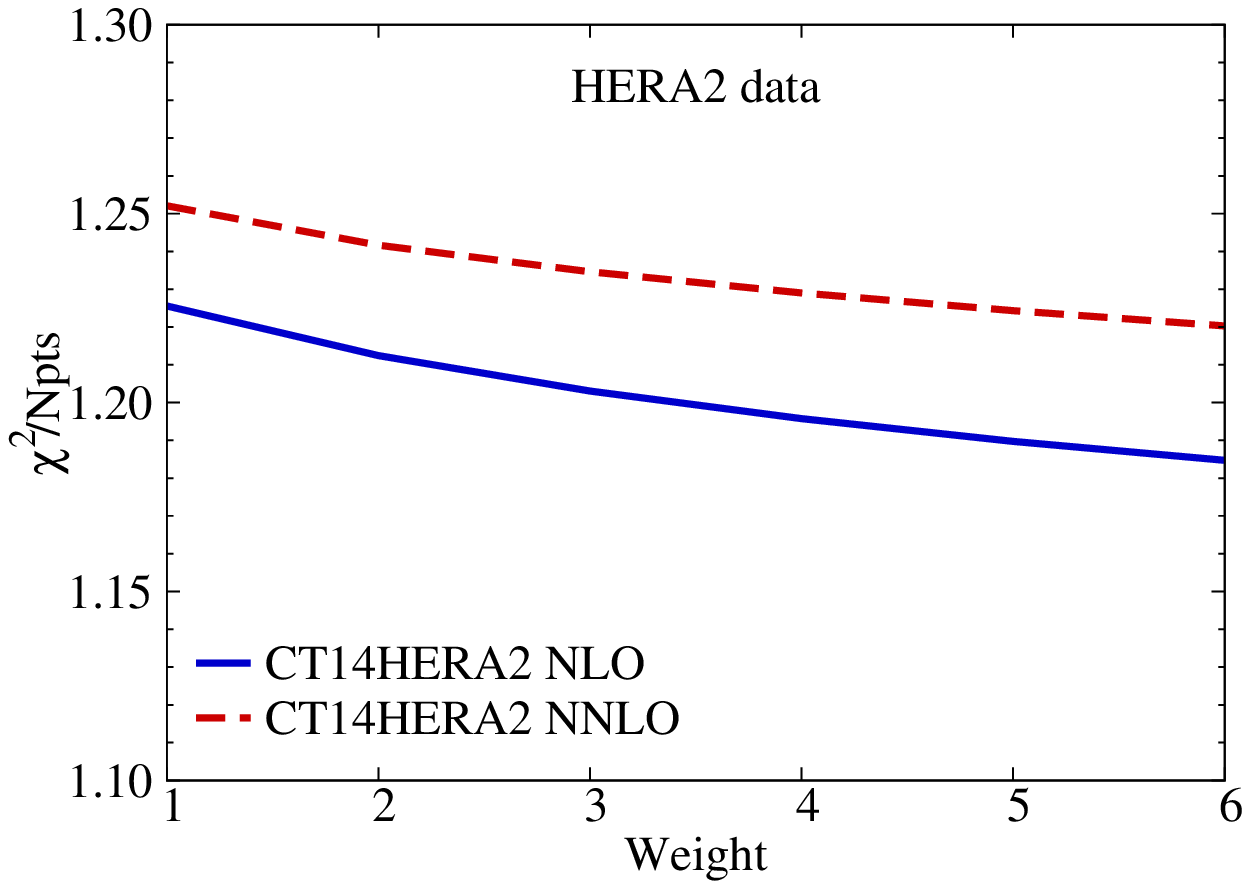}
\includegraphics[width=0.40\textwidth]
{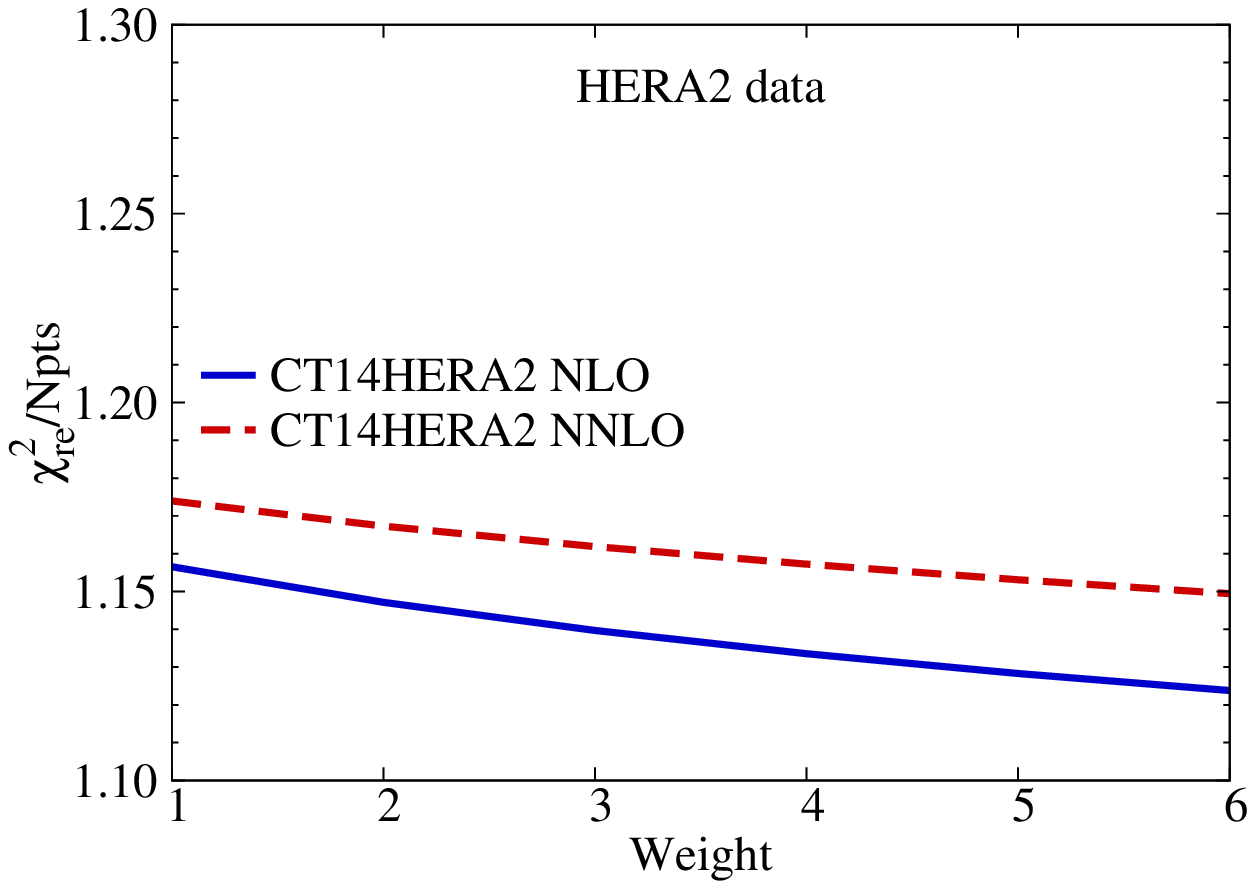}
\includegraphics[width=0.40\textwidth]
{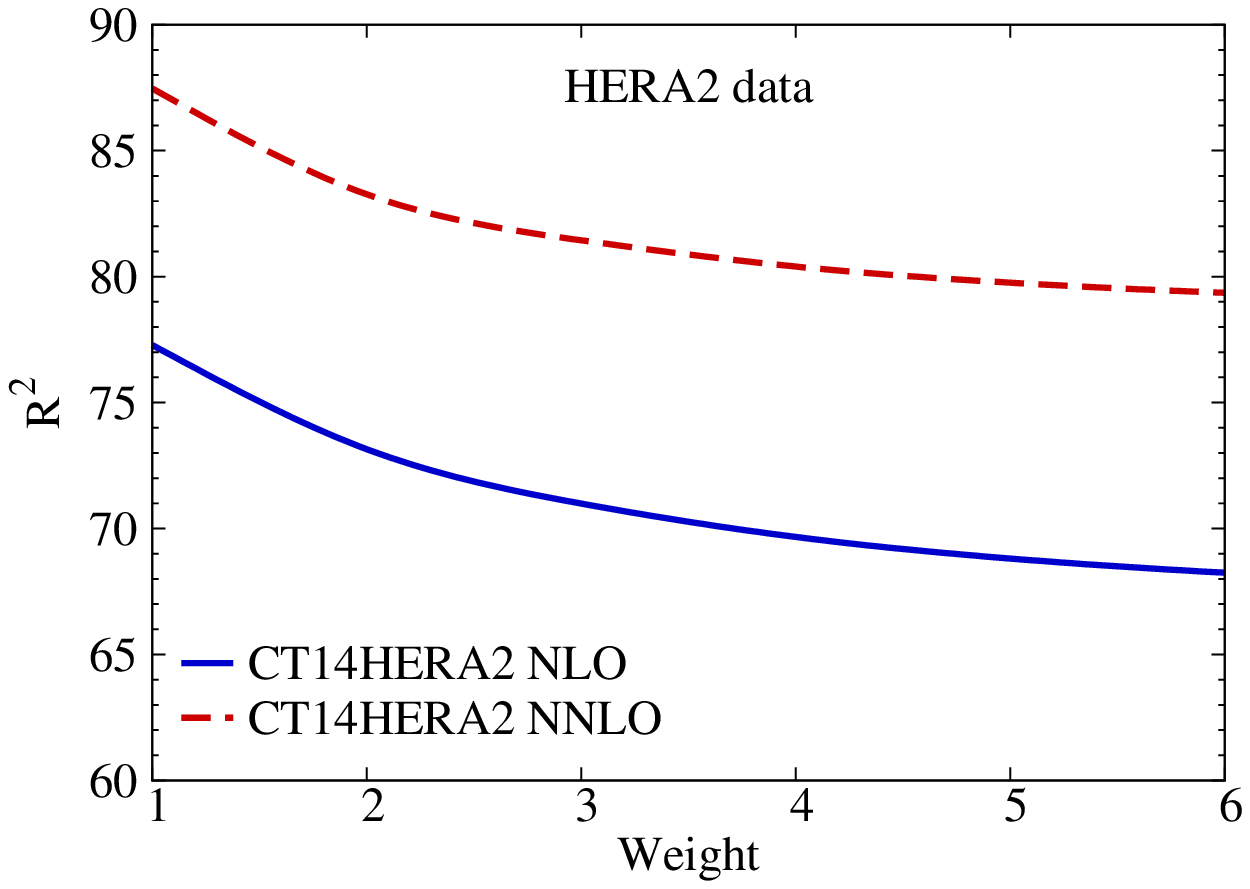}
\caption{Dependence of $\chi^2/N_{pts}$ (upper left),
$\chi^2_{re}/N_{pts}$ (upper right),
and $R^{2}$ penalty (lower panel) for HERA2 data
on the statistical weight assigned to the HERA2 data ensemble;
the PDFs are refitted for each weight.
\label{fig:CHIvWGT}}
\end{figure}

First, we examine how increasing the weight of HERA2 data
reduces $\chi^{2}/N_{pts}$ for the HERA2 data.
Figure~\ref{fig:CHIvWGT} shows $\chi^{2}/N_{pts}$ for the HERA2 combined data
($N_{pts} = 1120$) with \ZcthZ-like fits generated with weight factor varying  from 1 to 6,
at both NLO and NNLO accuracy.
The upper-left plot shows $\chi^{2}/N_{pts}$;
the upper-right plot shows $\chi^{2}_{\rm re}/N_{pts}$; and the lower
one shows $R^2$, the sum of the quadratic penalties
on the optimized systematic shifts in our treatment of correlated
systematic errors as nuisance parameters~\cite{Lai:2010vv}.
Of course, increasing the weight of the HERA2
data must cause $\chi^{2}/N_{pts}$ to decrease for that data.
But the change of $\chi^{2}$ is not large---only about $-5\%$
for a factor of 6 extra weighting.
The results are similar for NLO and NNLO.

\begin{figure}
\includegraphics[width=0.49\textwidth]
{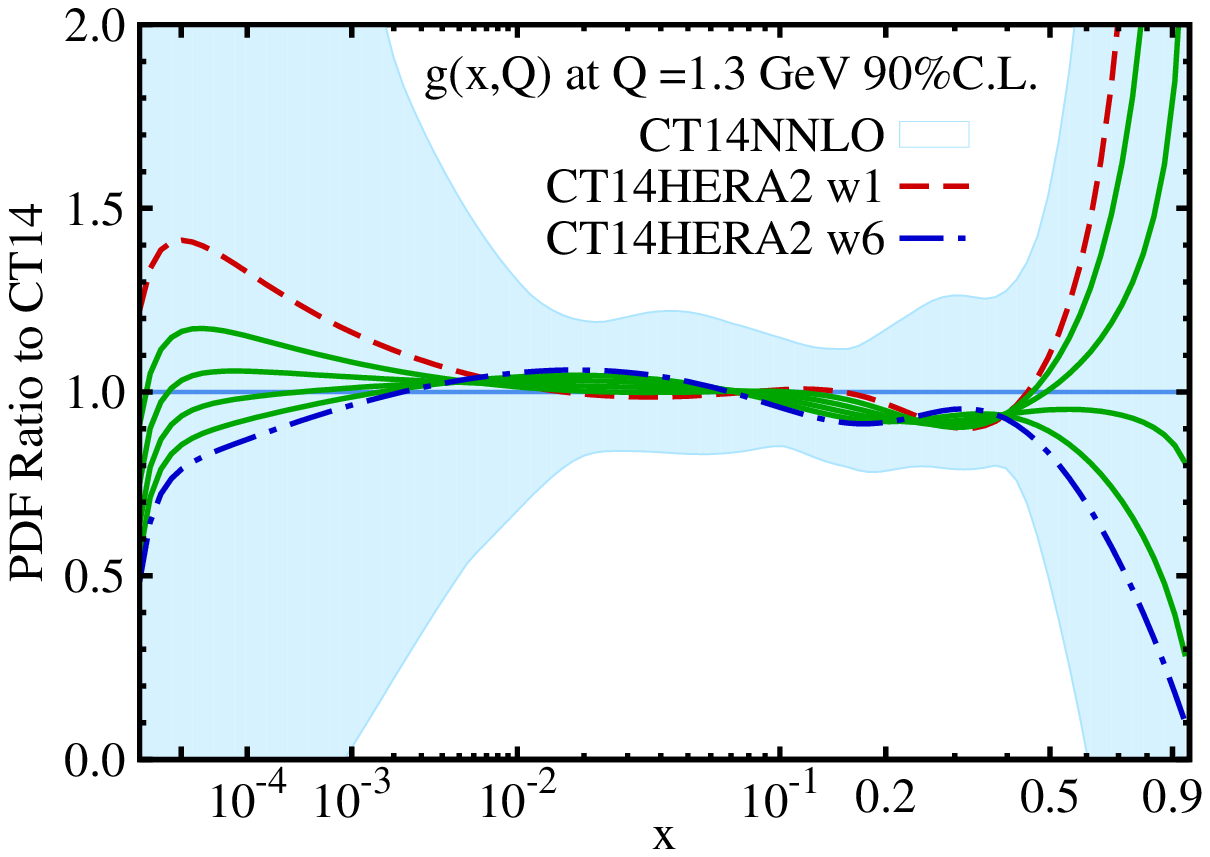}
\includegraphics[width=0.49\textwidth]
{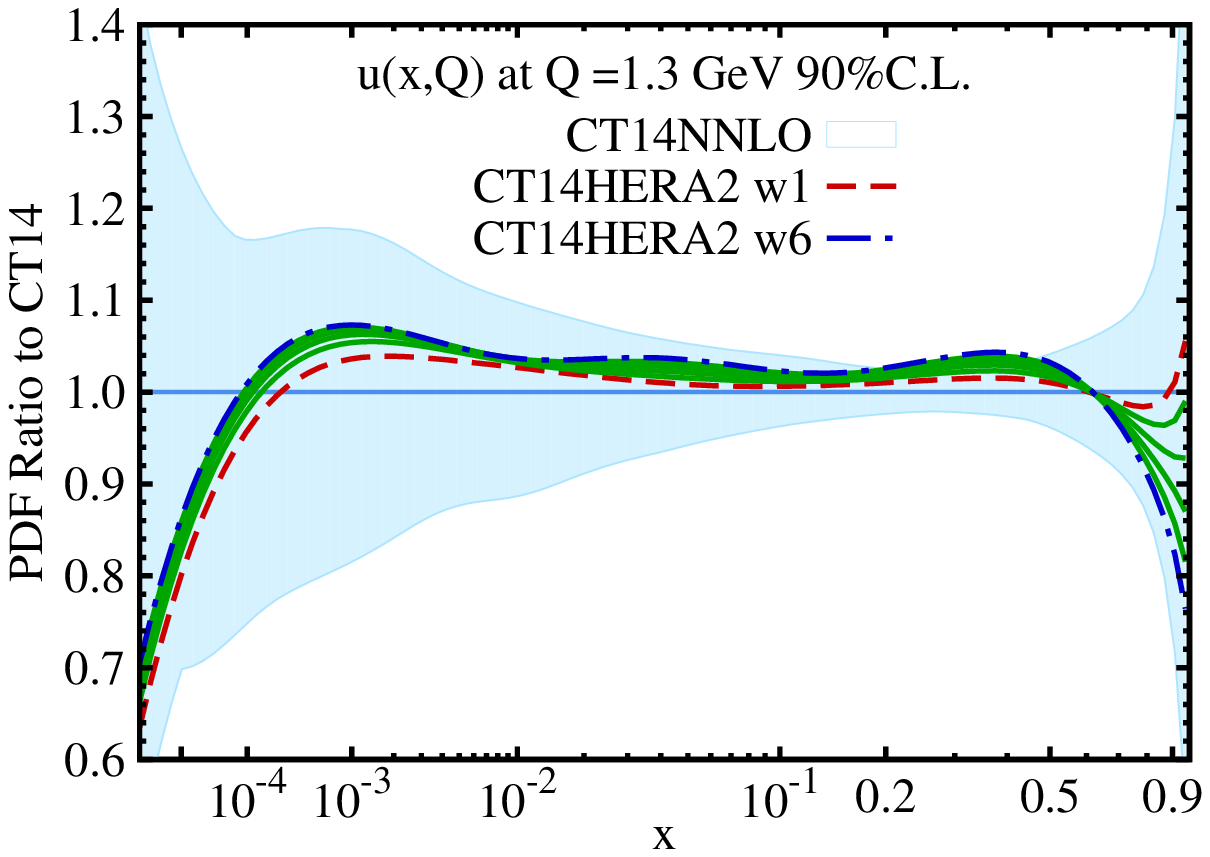}
\includegraphics[width=0.49\textwidth]
{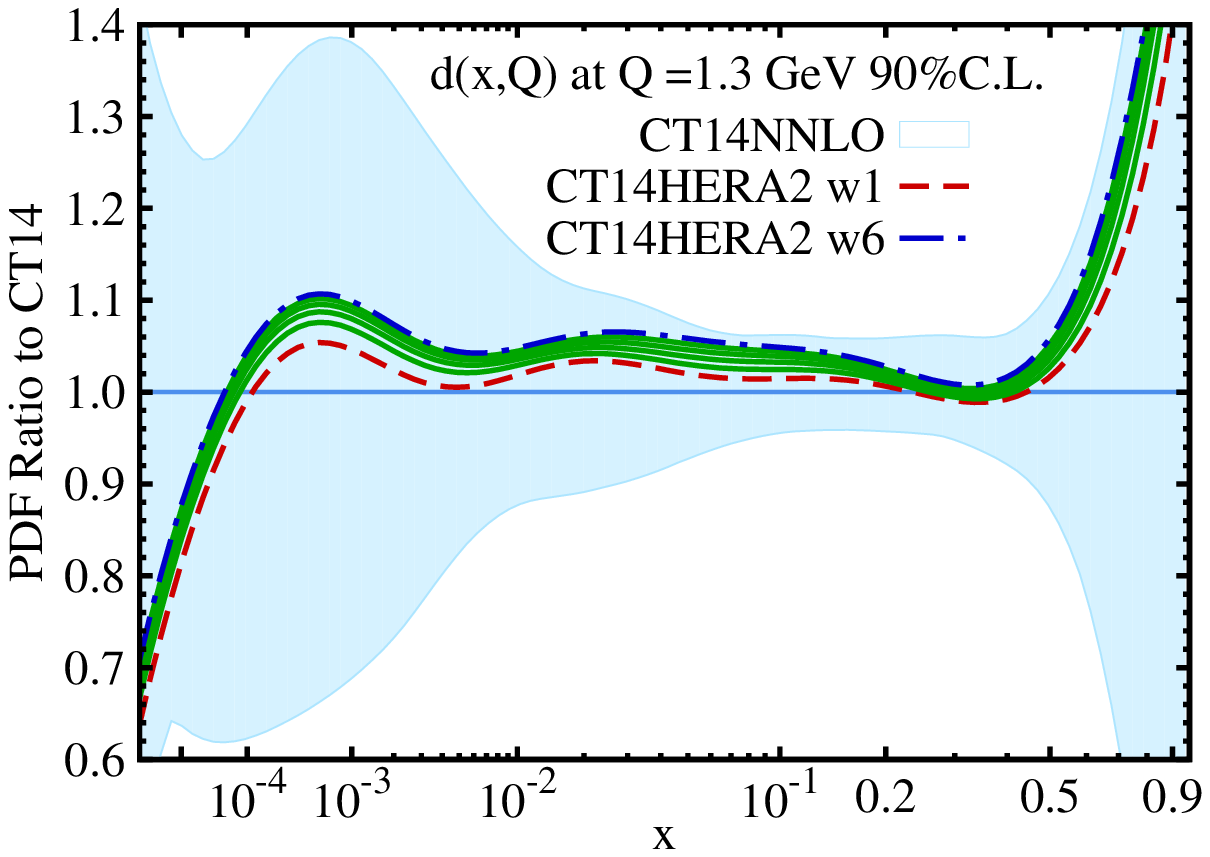}
\includegraphics[width=0.49\textwidth]
{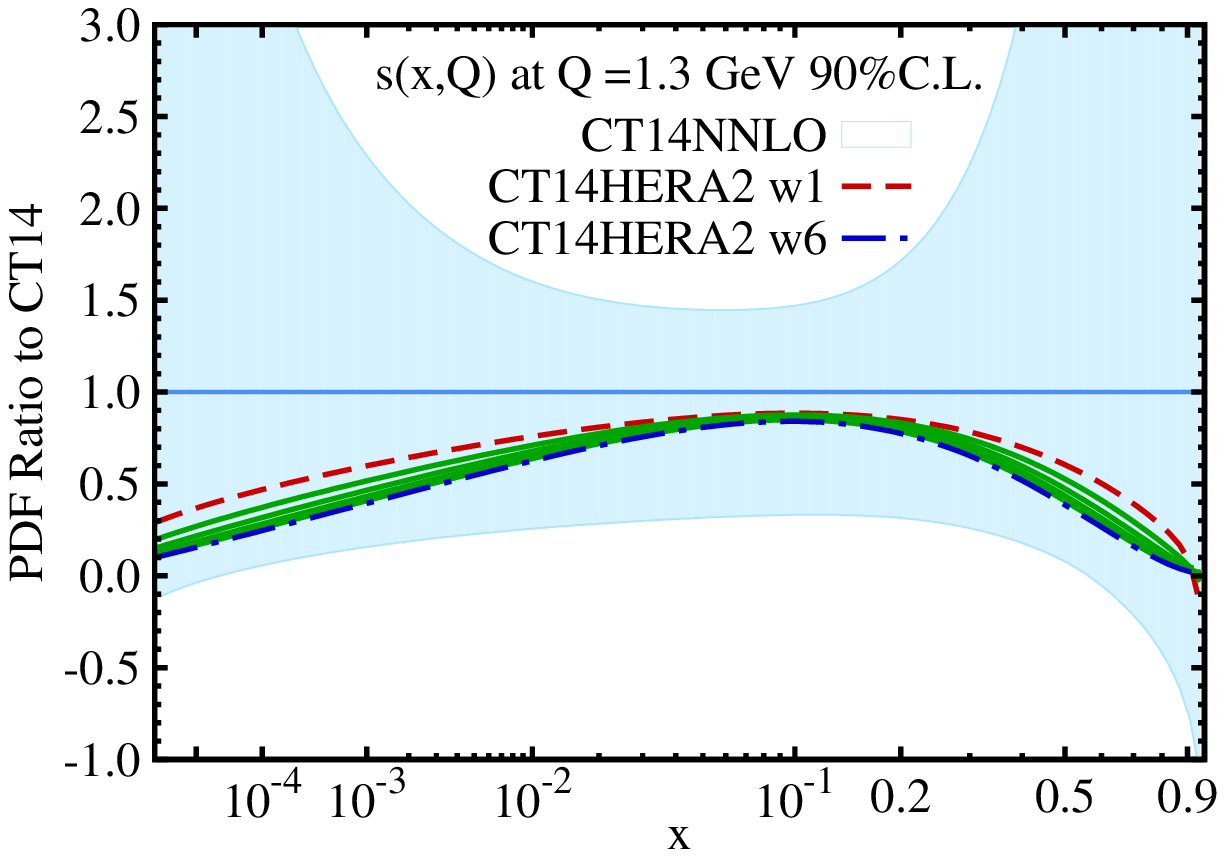}
\includegraphics[width=0.49\textwidth]
{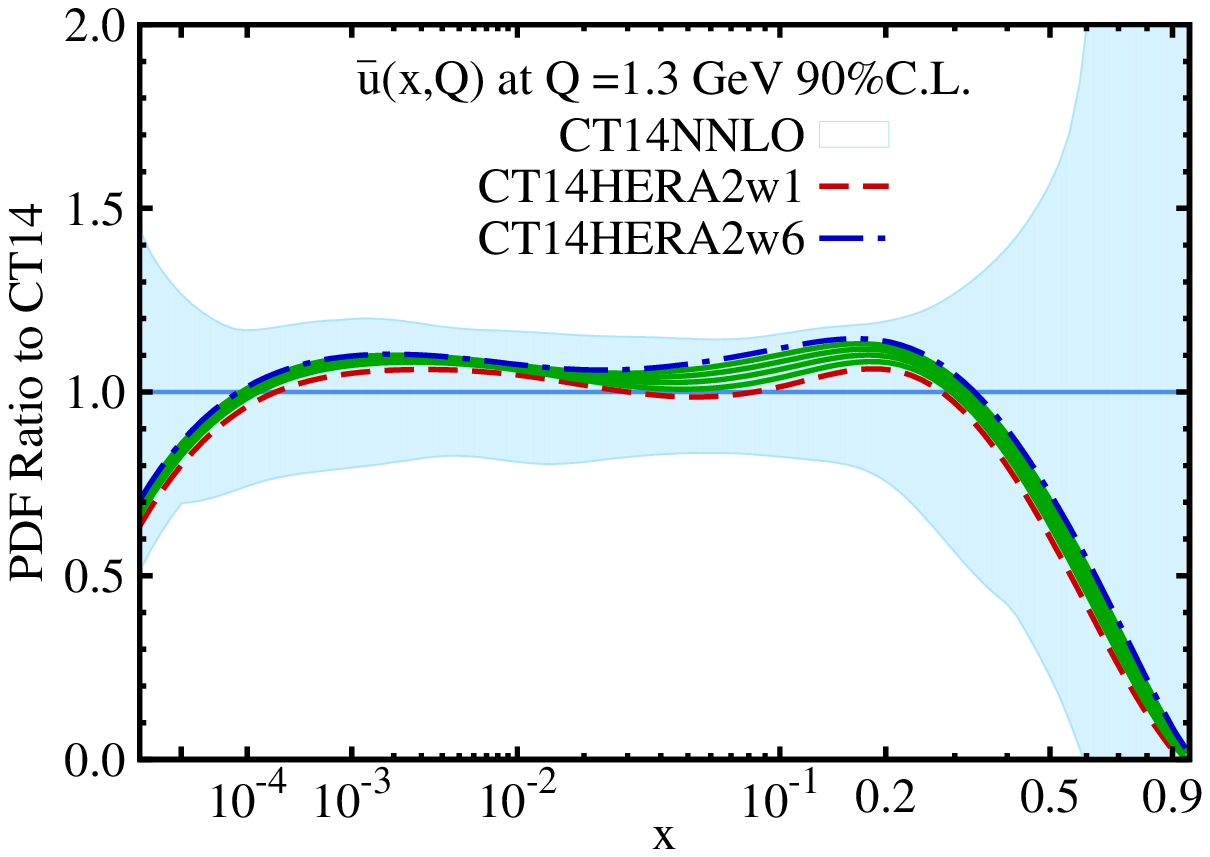}
\includegraphics[width=0.49\textwidth]
{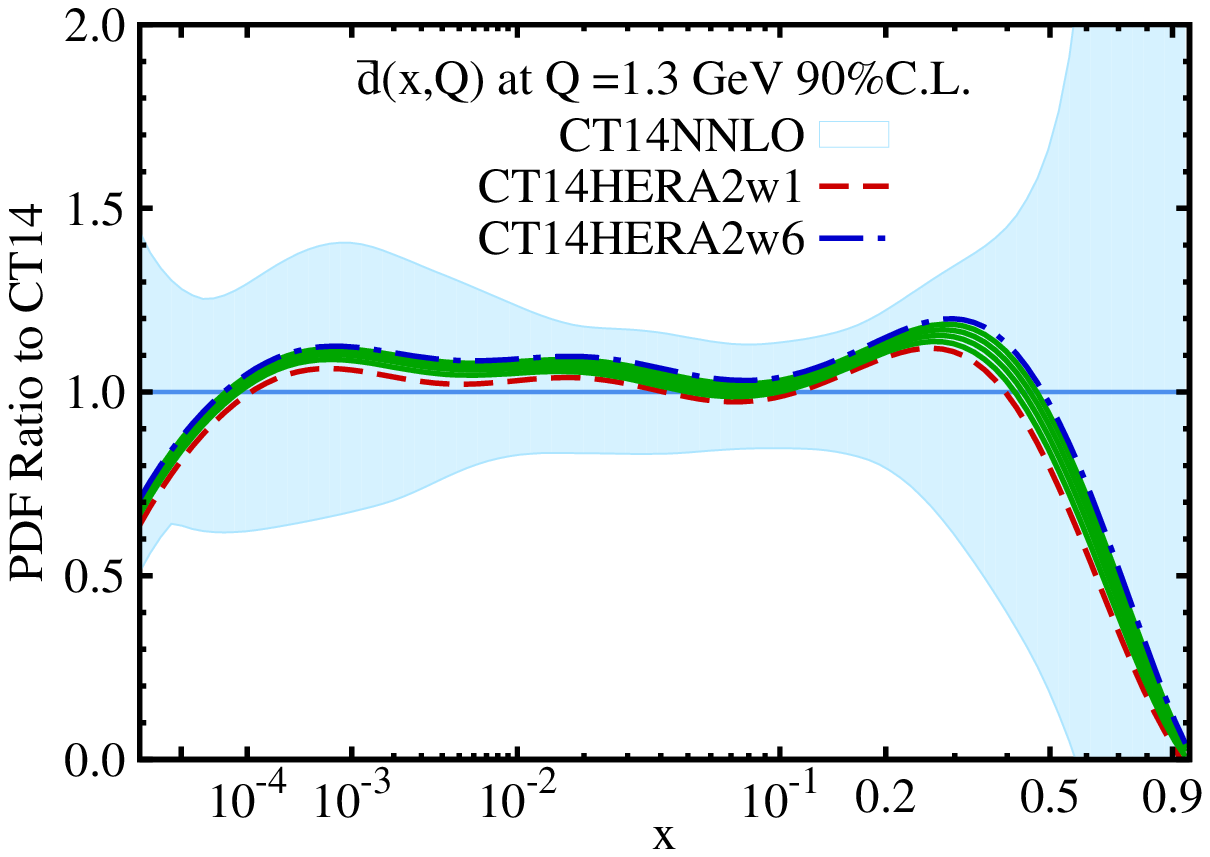}
\caption{
Comparison of \ZcthZ PDFs at $Q=1.3$ GeV within the CT14(NNLO)
uncertainty band.
Each curve represents the ratio of CT14$_\textrm{HERA2}$ / CT14
for a particular value of the weight assigned to the HERA2 data
in the global analysis.
The weight factors vary from 1 to 6.
\label{fig:PDFvWGT}}
\end{figure}

Secondly, as the weight of the HERA2 data set is increased, the resulting PDFs change, too.
Figure~\ref{fig:PDFvWGT} illustrates this,
by plotting the ratio of the CT14$_\textrm{HERA2}$ PDF to the CT14 PDF,
as a function of the weight factor assigned to the HERA2 data.
The HERA2 weights range from 1 to 6.
The uncertainty band of the CT14 PDF is also shown, evaluated at the
90\% confidence level (C.L.).
All PDFs are plotted at $Q = 1.3$ GeV.
For the gluon,
as the HERA2 weight increases, the CT14$_\textrm{HERA2}$ PDF decreases
at $x\lesssim 10^{-3}$ and decreases rapidly at $x > 0.4$;
for intermediate $x$ values,
$g(x,Q_{0})$ varies by a few percent.
For the up quark,
the PDF exhibits a modest fractional increase in the central $x$ region
(for $0.01 < x < 0.5$) relative to its PDF error band,
as the HERA2 weight increases.
The down quark PDF has a similar behavior for $0.01 < x < 0.5$
but with larger magnitude of variation than the up quark.
Similarly, for the up antiquark, the PDF exhibits a modest fractional increase
for $x$ around $0.1$ to $0.2$, as the HERA2 weight increases;
and the down antiquark PDF has a similar increase for $x$ around $0.3$.
In contrast to the up and down flavors,
the strange quark PDF is reduced relative to CT14.
The reduction of $s(x,Q_{0})$
is mainly caused by freeing the parameter $a_{1}(s)$.
But, as we weight the HERA2 data more heavily,
$s(x,Q_{0})$ decreases even further.
We note that the same conclusion also holds for the CT14 NLO PDFs.

\section{Impact of data selection cuts on the fit to HERA2 Data}
\label{kinematic-cuts}

The HERA2 publication~\cite{Abramowicz:2015mha}
found that both HERAPDF2.0 PDFs and $\chi^2$ values
depend significantly on the choice of $Q_{\rm cut}$,
the minimum value of the four-momentum-transfer $Q$ in the HERA2 analysis.
In this section we explore the impact of variations of $Q_{\rm cut}$ on the \ZcthZ global analysis.

We perform multiple fits of \ZcthZ PDFs,
in which $Q_{\rm cut}$ is varied from 2 to 6 GeV, and
compare the results to the previous findings of the CT14 analysis.
For every choice of $Q_{\rm cut}$, we report
the total $\chi^{2}$, reduced $\chi^2$ ({\it i.e.}, $\chi^2_{\rm re}$),
and systematic shift
penalty $R^2$ defined by Eq.~(\ref{chi2}), together with
the number of data points $N_{pts}$ in parentheses.
Tables~\ref{tbl:subprocHERA1} and \ref{tbl:subprocHERA2} show these
quantities for the HERA1 and HERA2 data, compared to the theoretical
predictions based on CT14 NNLO and \ZcthZ NNLO PDFs, respectively. The lower
parts of each table show the breakdown of
$\chi^2_{\rm re}$ and numbers of points over the four contributing
DIS subprocesses, in NC and CC
interactions: NC $e^{+}p$,  NC $e^{-}p$, CC $e^{+}p$, and CC $e^{-}p$.

In the CT14 analysis
the subsets of HERA1 data have small values of
$\chi^2_{\rm re}/N_{pts}$, as shown in Table~\ref{tbl:subprocHERA1}.
For the $e^{-}p$ processes, $\chi^{2}_{\rm re}/N_{pts}$ is less than $1$;
for the $e^{+}p$ processes, $\chi^{2}_{\rm re}/N_{pts}$ is approximately $1$.
Also, there is no dependence on $Q_{\rm cut}$, except for a small
decrease in $\chi^{2}_{\rm re}/N_{pts}$ for the case of NC $e^{+}p$.
The {\em total} $\chi^{2}/N_{pts}$ decreases with $Q_{\rm cut}$ because
the NC $e^{+}p$ subset dominates the total.
We conclude that, for the CT14/HERA1 analysis,
the standard choice $Q_{\rm cut} = 2$ GeV is not
qualitatively different from the other $Q_{cut}$ choices
in the 2 to 6 GeV range.

\begin{table}[ht]
\begin{tabular}{|l|c|c|c|c|c|c|c|c|c|c|c|c|c|c|}
\hline
$Q_{\textrm{cut}}$ [GeV] &
 No cut   & 2.00       & 3.87       & 4.69       & 5.90       \tabularnewline
\hline
$\chi^2/N_{pts}(N_{pts})$ &
 (647)    & 1.02 (579) & 0.93 (516) & 0.93 (493) & 0.91 (470) \tabularnewline
\hline
$R^2/114(R^2)$ &
          & 0.43(48.80)& 0.24(27.34)& 0.25(28.38)& 0.25(28.48)\tabularnewline
\hline
$\chi^2_{\rm re}/N_{pts}(N_{pts})$ &
 (647)    & 0.94 (579) & 0.89 (516) & 0.87 (493) & 0.84 (470) \tabularnewline
\hline
\hline
NC $e^+ p$ &
 (434)    & 1.05 (366) & 0.96 (303) & 0.96 (280) & 0.92 (257) \tabularnewline
\hline
NC $e^- p$ &
 (145)    & 0.74 (145) & 0.75 (145) & 0.75 (145) & 0.75 (145) \tabularnewline
\hline
CC $e^+ p$ &
 (34)     & 0.97 (34)  & 0.98 (34)  & 0.99 (34)  & 0.99 (34)  \tabularnewline
\hline
CC $e^- p$ &
 (34)     & 0.53 (34)  & 0.53 (34)  & 0.53 (34)  & 0.53 (34)  \tabularnewline
\hline
\end{tabular}
\caption{Goodness-of-fit characteristics for the HERA1 combined data
  with specified $Q_{cut}$ selection constraints,
  and theory predictions based on the CT14 NNLO PDFs
  determined with the nominal cut $Q_{cut} \geq 2$ GeV.
The four lowest rows give $\chi^{2}_{\rm re}/N_{pts}$ for each DIS subprocess.
\label{tbl:subprocHERA1}}
\end{table}

In the \ZcthZ/HERA2 analysis (Table~\ref{tbl:subprocHERA2}),
the values of $\chi^{2}_{\rm re}/N_{pts}$ are larger than 1
for the subprocesses,
and much larger in the cases of $e^{-}p$ scattering.
The PDFs for the different columns of Table III
were refitted for each choice of $Q_{\rm cut}$.
Even with the refitting,
the values of $\chi^{2}_{\rm re}/N_{pts}$ remain large.
The dependence of $\chi^{2}_{\rm re}/N_{pts}$ on $Q_{\rm cut}$
is small for NC $e^{+}p$ and negligible for the other three cases.

\begin{table}[ht]
\begin{tabular}{|l|c|c|c|c|c|c|c|c|c|c|c|c|c|}
\hline
$Q_{\textrm{cut}}$ [GeV] &
 No cut   & 2.00         & 3.87       & 4.69        & 5.90       \tabularnewline
\hline
$\chi^2/N_{pts}(N_{pts})$ &
 (1306)   &  1.25 (1120) & 1.19 (967) & 1.21 (882)  & 1.23 (842) \tabularnewline
\hline
$R^2/170(R^2)$ &
          &  0.51 (87.47)& 0.29(49.11)& 0.29 (48.99)& 0.29 (49.40)\tabularnewline
\hline
$\chi^2_{\rm re}/N_{pts}(N_{pts})$ &
 (1306)   &  1.17 (1120) & 1.14 (967) & 1.15 (882)  & 1.18 (842) \tabularnewline
\hline
\hline
NC $e^+p$ &
 (1066)   &  1.11 (880)  & 1.06 (727) & 1.06 (642)  & 1.09 (602) \tabularnewline
\hline
NC $e^-p$ &
 (159)    &  1.45 (159)  & 1.44 (159) & 1.45 (159)  & 1.45 (159) \tabularnewline
\hline
CC $e^+p$ &
 (39)     &  1.10 (39)   & 1.10 (39)  & 1.10 (39)   & 1.10 (39)  \tabularnewline
\hline
CC $e^-p$ &
  (42)    &  1.52 (42)   & 1.50 (42)  & 1.50 (42)   & 1.50 (42)  \tabularnewline
\hline
\end{tabular}
\caption{Goodness-of-fit characteristics for the HERA2 combined data
  with specified $Q_{cut}$ selection constraints,
  and theory predictions based on the \ZcthZ NNLO PDFs
  refitted with the same $Q_{cut}$ value.
\label{tbl:subprocHERA2}}
\end{table}

In contrast to CT14, in the \ZcthZ analysis
we see only small variations in $\chi^{2}_{\rm re}/N_{pts}$
with the four values of $Q_{\rm cut}$.
We note that the apparent large change in $\chi^{2}/N_{pts}$
from $Q_{\rm cut}$ of 2 to 3.87\,GeV, as shown in the second row of
Table \ref{tbl:subprocHERA2}, is due to the change in $R^2$ values in
the third row. Recall that $\chi^2$  is given by the sum of
  $\chi^2_{re}$, which changes little, and $R^2$, which decreases
  from 2 GeV
  to 3.87 GeV.
With a larger $Q_{\rm cut}$ value, at 3.87\,GeV, there are fewer data points to be fit with the same number of correlated
systematic errors (170 in the \ZcthZ analysis), hence it leads to
a smaller $R^2/170$ value, from 0.51 to 0.29.

\begin{figure}[htbp]
\includegraphics[width=0.49\textwidth]
{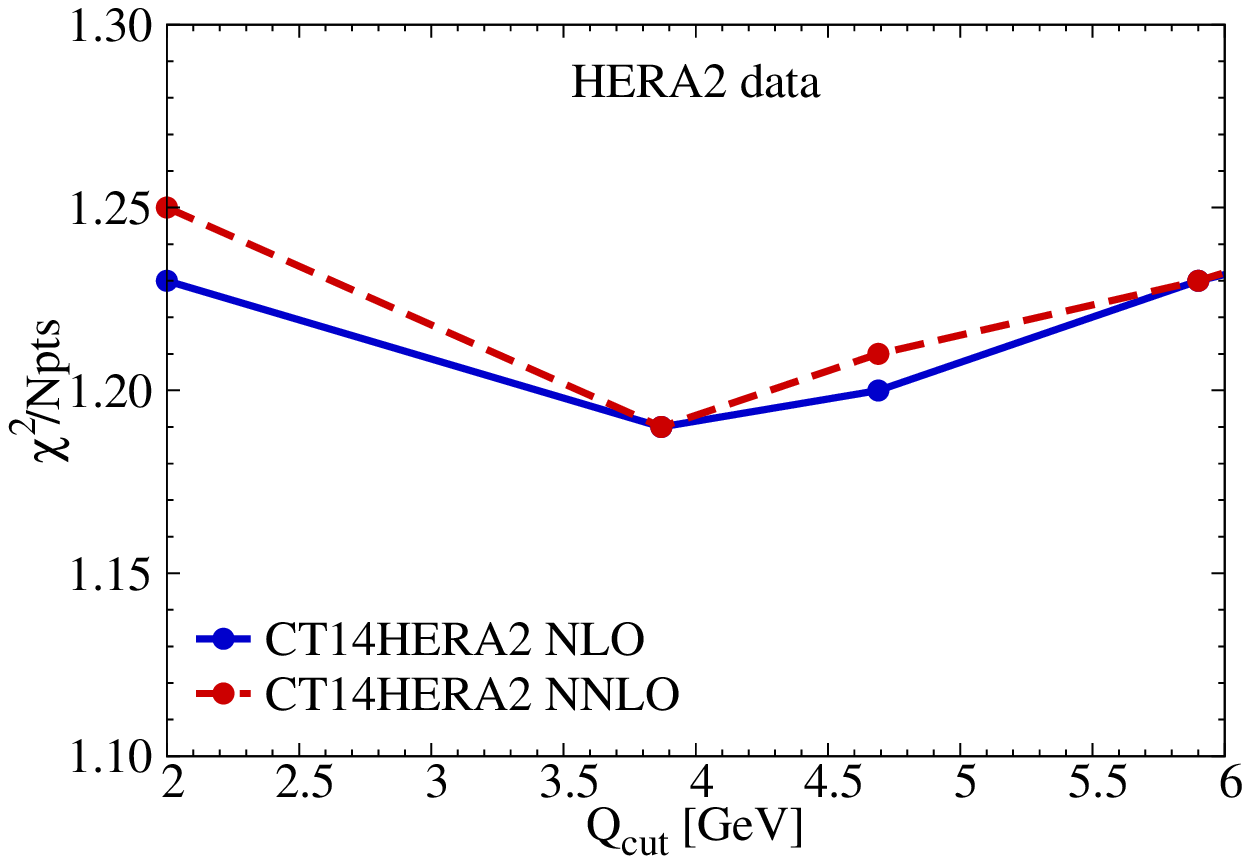}
\includegraphics[width=0.49\textwidth]
{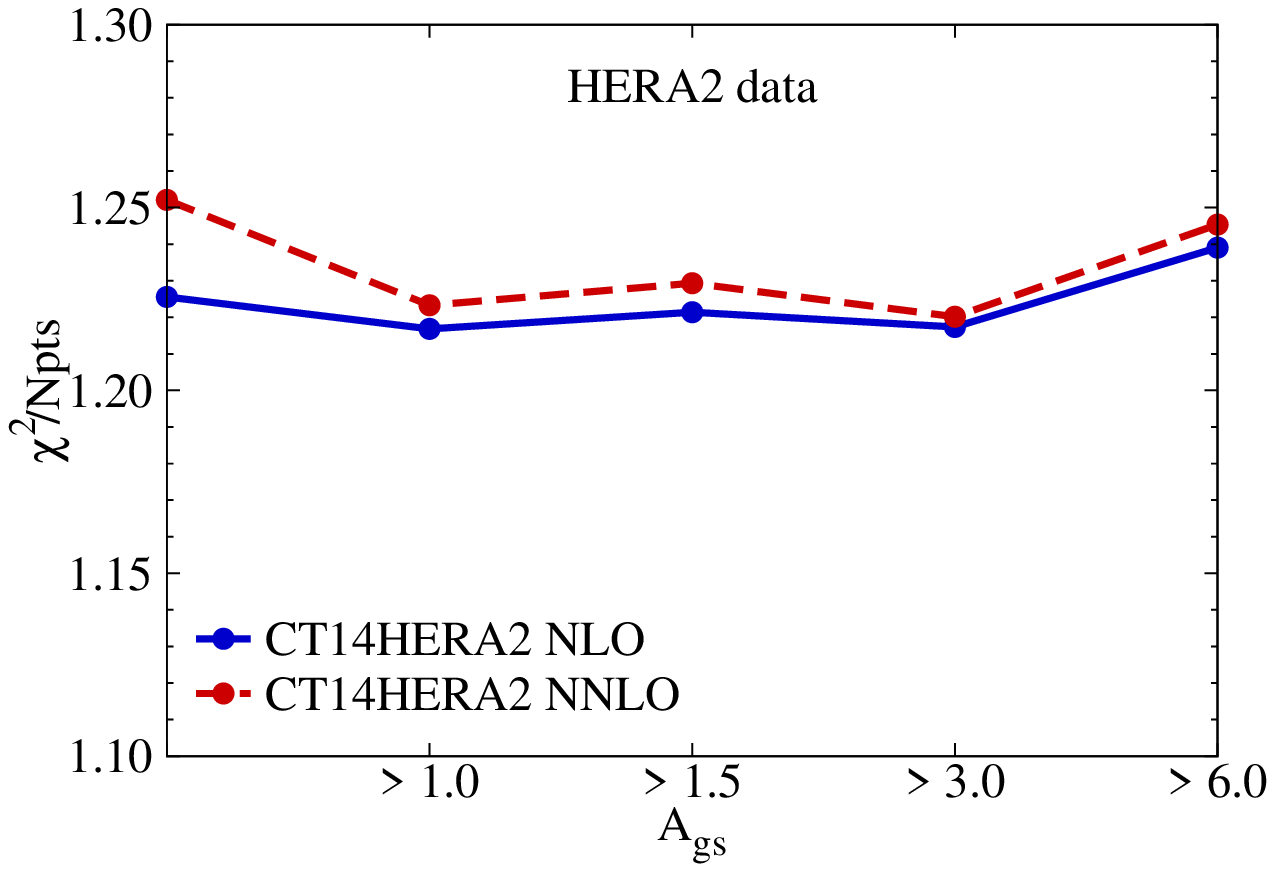}
\includegraphics[width=0.49\textwidth]
{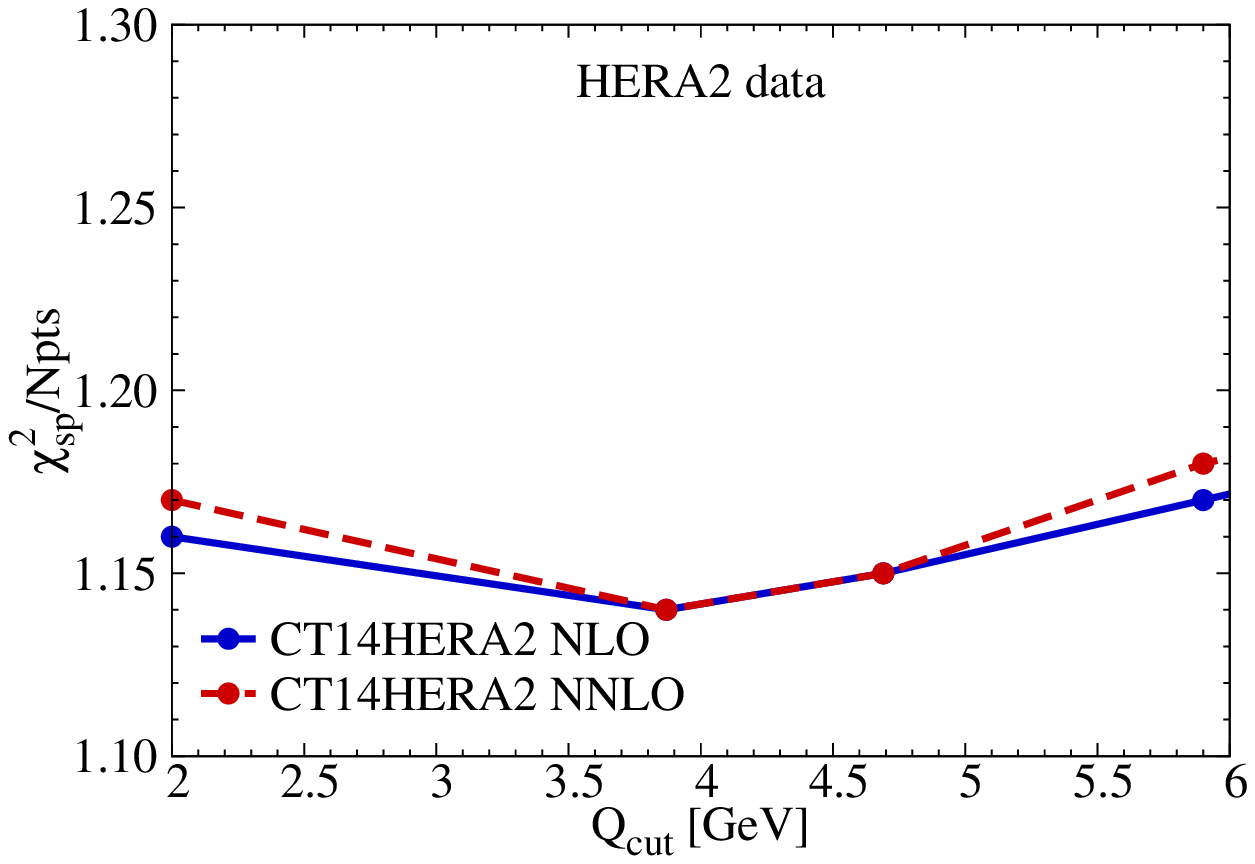}
\includegraphics[width=0.49\textwidth]
{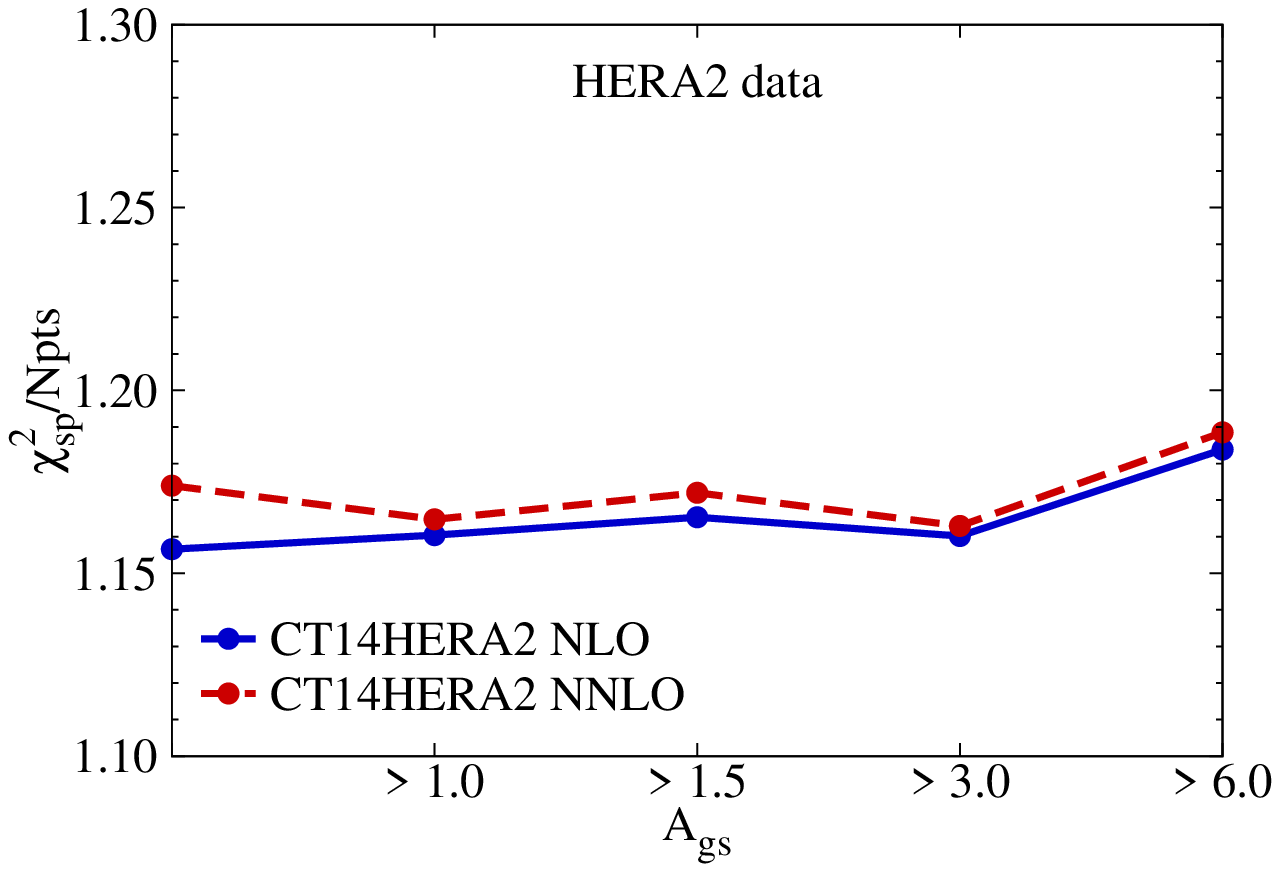}
\includegraphics[width=0.49\textwidth]
{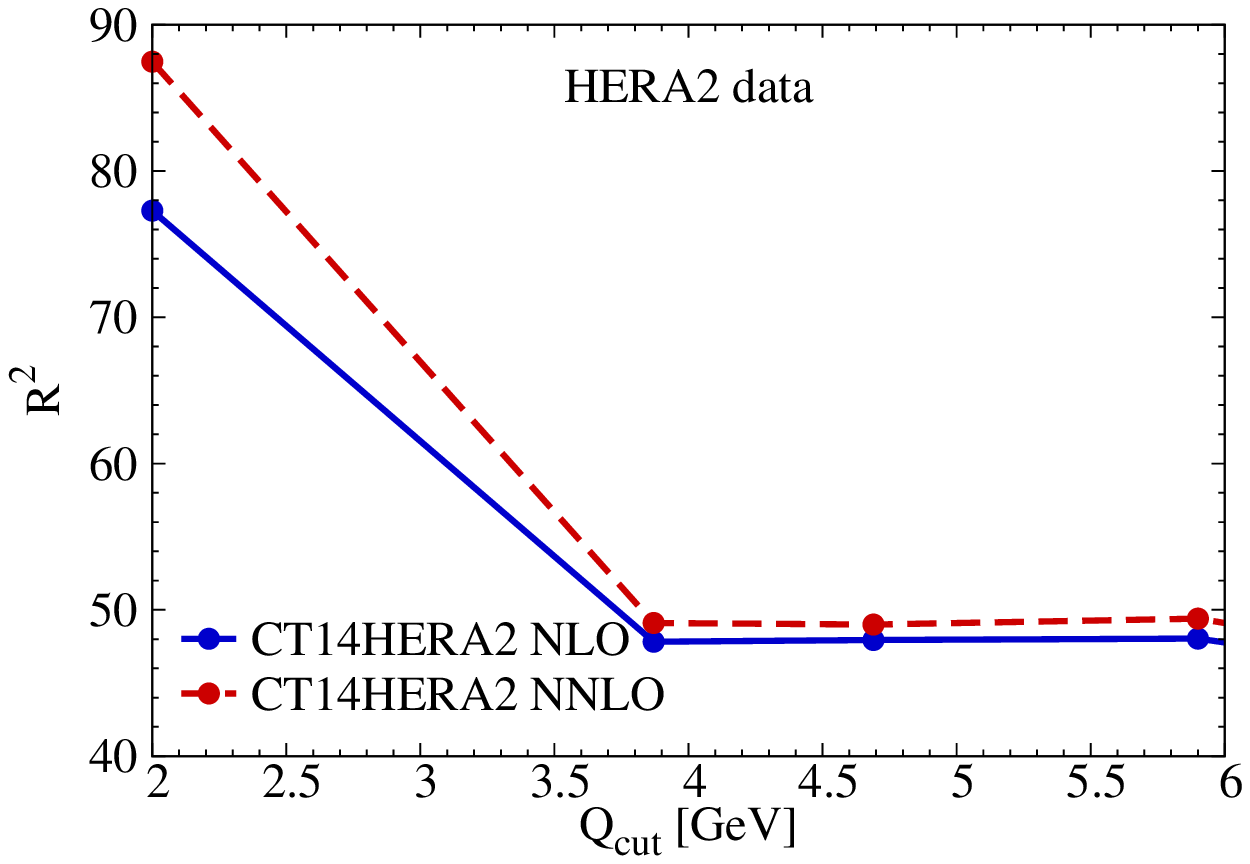}
\includegraphics[width=0.49\textwidth]
{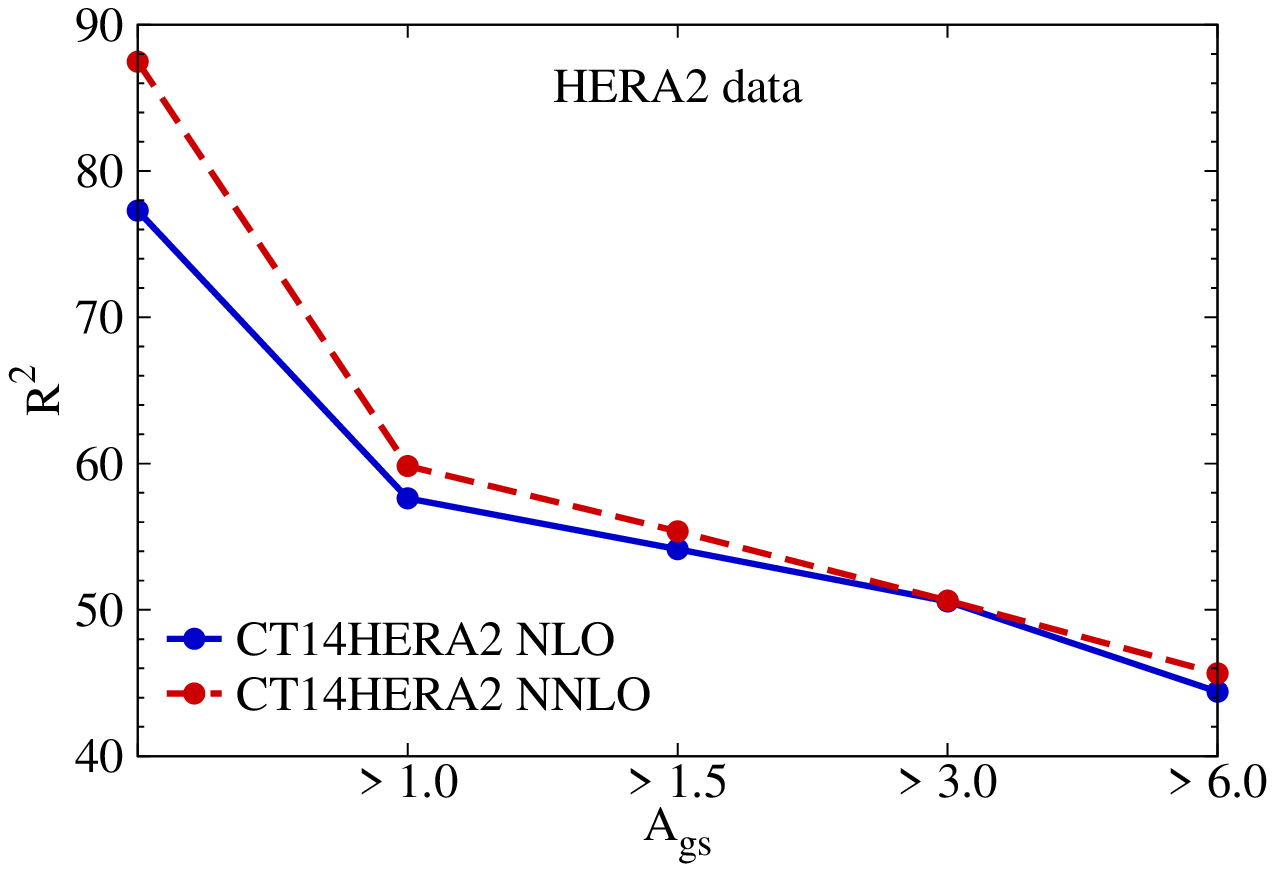}
\caption{
Left panels:
$\chi^{2}/N_{pts}$ (top), reduced $\chi^{2}/N_{pts}$ (middle),
and $R^{2}$ (bottom)
for the HERA2 data and \ZcthZ PDFs,
as a function of $Q_{\rm cut}$.
Right panels:
The same as a function of the cutoff value
of the geometric scaling variable $A_{\rm gs}$.
\label{fig:cut1}}
\end{figure}

Figure~\ref{fig:cut1} shows the results
on $\chi^{2}$ versus $Q_{\rm cut}$ of Table \ref{tbl:subprocHERA2}
in graphical form.
The behavior of $\chi^2/N_{pts}$ for the HERA2 data
(sum of all four subprocesses)
is illustrated in the left panels of Fig.~\ref{fig:cut1}.
The graphs show the dependence on $Q_{\textrm{cut}}$
in the \ZcthZ analysis at both NLO and NNLO.
The upper panel is $\chi^{2}$ and the middle panel
is the reduced $\chi^{2}$, versus $Q_{\rm cut}$.
The values of $\chi^{2}/N_{pts}$ for the HERA2 data
exhibit a {\em shallow minimum} for $Q_{\rm cut}$
in the range
$3.5\,{\rm GeV} \lesssim Q_{\rm cut} \lesssim 4$ GeV.
The reduction of $\chi^{2}_{re}$ at $Q_{\rm cut} \sim 4$ GeV,
compared to our standard choice of $Q_{\rm cut}=2$\,GeV,
from 1.17 to 1.15, does not seem significant.
An interesting feature of the graphs is that
near the minimum the NNLO and NLO results are equal,
whereas NNLO has slightly larger $\chi^{2}$ on either side of the minimum.

The lower panel in Fig.\ \ref{fig:cut1} shows $R^{2}$,
the total quadratic penalty for the systematic errors,
as a function of $Q_{\rm cut}$.
The value of $R^2$ decreases significantly from $Q_{\rm cut}$ = 2 GeV
to 3.87 GeV, from 87 to 49.
For ideal Gaussian systematic errors we would expect $R^{2} \sim 170$
for 170 systematic errors.
When the low-$Q$ data points are discarded by the cut,
the systematic errors become less important.
However, this reduction of $R^{2}$ is shared by 1120 total data points,
so the overall net change in $\chi^{2}/N_{pts}$ is mild.

\setcounter{subsubsection}{0}
\subsubsection{Dependence on the geometric rescaling variable}

While Fig\ \ref{fig:cut1} examines dependence of fits on $Q$ cuts
that are imposed independently of the Bjorken $x$ value,
it is as instructive to consider the dependence of $\chi^2$
on correlated cuts in $Q$ and $x$.
For this purpose
we define the geometric scaling variable $A_{\rm gs} = x^{\lambda} Q^2$,
where $\lambda$ is a parameter set equal to $0.3$ in this
study~\cite{Stasto:2000er,Caola:2009iy,Lai:2010vv}.
The $A_{\rm gs}$ variable can be utilized to explore
the impact of data in kinematic regions of both
small $Q$ and small $x$.
We can test whether the goodness of fit
improves if we exclude data at small $\{x,Q\}$.
The variable $A_{\rm gs}$ has been used in previous analyses
to search for possible deviations from DGLAP evolution due to
saturation or small-$x$ related
phenomena~\cite{Stasto:2000er,Caola:2009iy}.
The basic method is to
(i) generate PDFs using data in the kinematic region
above the $A_{gs}$ cut in the $x$ and $Q$ plane,
where the NLO/NNLO DGLAP factorization
is supposed to be valid;
(ii) then use DGLAP evolution equations to evolve these PDFs down to the
low-$x$ and $Q$ region below the $A_{gs}$ cut,
where one might expect possible deviations;
(iii) finally, compare predictions to the data in the low $A_{gs}$
region, which was not used for PDF determination.
The portion of HERA2 data that is excluded by varying
$(A_{\rm gs})_{\rm cut}$
from 1.0 to 6.0 is shown in Fig.~\ref{chi2res} (the lower right inset).
The results of the fits for various choices of
$(A_{\rm gs})_{\rm cut} $,
at both NLO and NNLO accuracy,
are illustrated in the right panels of Fig.~\ref{fig:cut1}.
(The upper panel is $\chi^{2}$,
the middle panel is reduced $\chi^{2}$,
and the lower panel is $R^{2}$.)
The values of $\chi^{2}/N_{pts}$ for four choices of
$(A_{\rm gs})_{\rm cut} $ are shown.
Here, we consider only data points with $Q$ values greater than 2 GeV
in order to validate the application of the perturbative DGLAP evolution
equation. We find that the behavior of $\chi^2$ has small variations,
and they are not monotonic. Hence, we conclude that our analysis
of HERA2 data does not indicate clear deviations from DGLAP
evolution. Alternatively, one could include also the data points
below the $A_{gs}$ cut (though still with $Q > 2$ GeV)
in the calculation of $\chi^2$ in the final comparison while
fitting only the data above the $A_{gs}$ cut.
We found a similar conclusion as that carried out for the CT10 NLO PDFs,
as shown in the appendix of Ref.~\cite{Lai:2010vv}.
For example, the value of $\chi^{2}_{res}/N_{pts}$ of the combined
HERA2 data set, with $A_{\rm gs} > 1.5$, increases by about $0.2-0.3$
units as compared to that without any $A_{\rm gs}$ cut.
This result does not change much even when we use a more
flexible gluon PDF, by introducing one more nonperturbative
shape parameter, in the fit.
Furthermore, the value of $\chi^{2}_{res}/N_{pts}$ for the NLO fit
is larger than the NNLO fit by about 0.1 unit, which is about the same
size as the variation from including the $A_{\rm gs} > 1.5$ cut in the fit.
This is comparable with the usual uncertainties and consistent with
the above conclusion that the HERA2 data do not show clear deviations
from DGLAP evolution.


\section{Comparison of CT14$_\textrm{HERA2}$ and CT14 PDFs}

In this section we describe the changes in central values
and uncertainties of CT14$_\textrm{HERA2}$ PDFs, which are
obtained from our global analysis with the weight
of HERA2 data set to be 1,
compared to CT14 PDFs.
Here, $Q$ is equal to the initial scale $Q_0=1.3$ GeV;
also, only the NNLO PDFs are shown.
At this low scale, the PDF
uncertainties are magnified, and they are reduced at electroweak
scales as a consequence of DGLAP evolution.
Additional plots can be found on the CTEQ public website
\cite{cteqweb}.

\begin{figure}
\includegraphics[width=0.47\textwidth]
{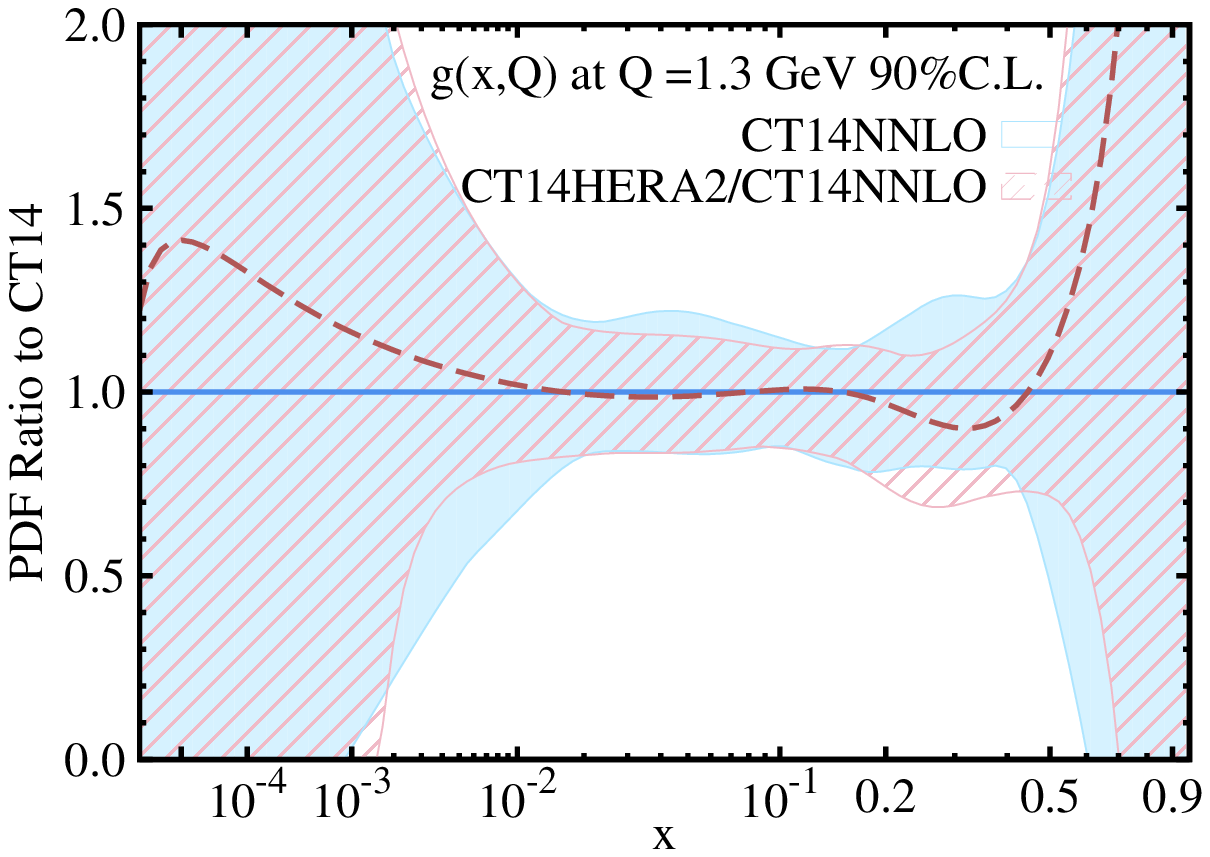}
\includegraphics[width=0.47\textwidth]
{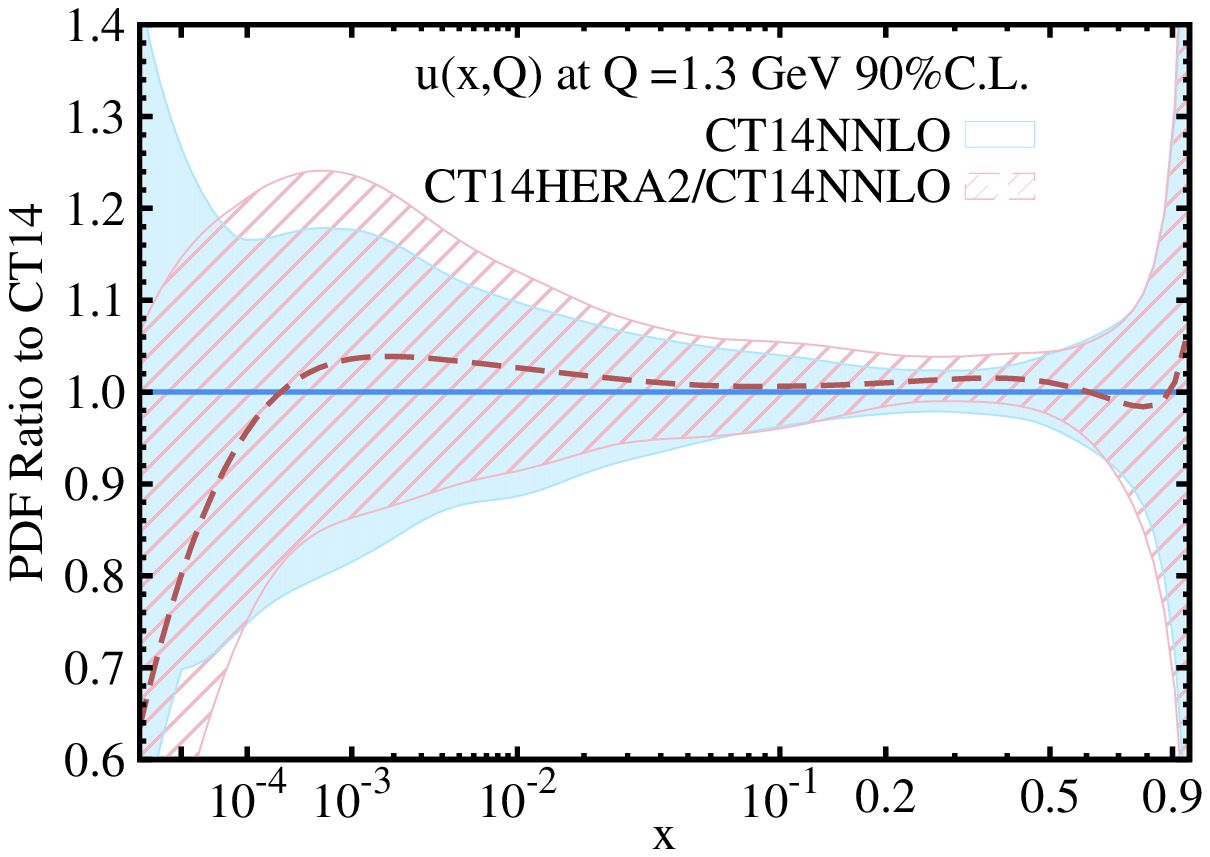}
\includegraphics[width=0.47\textwidth]
{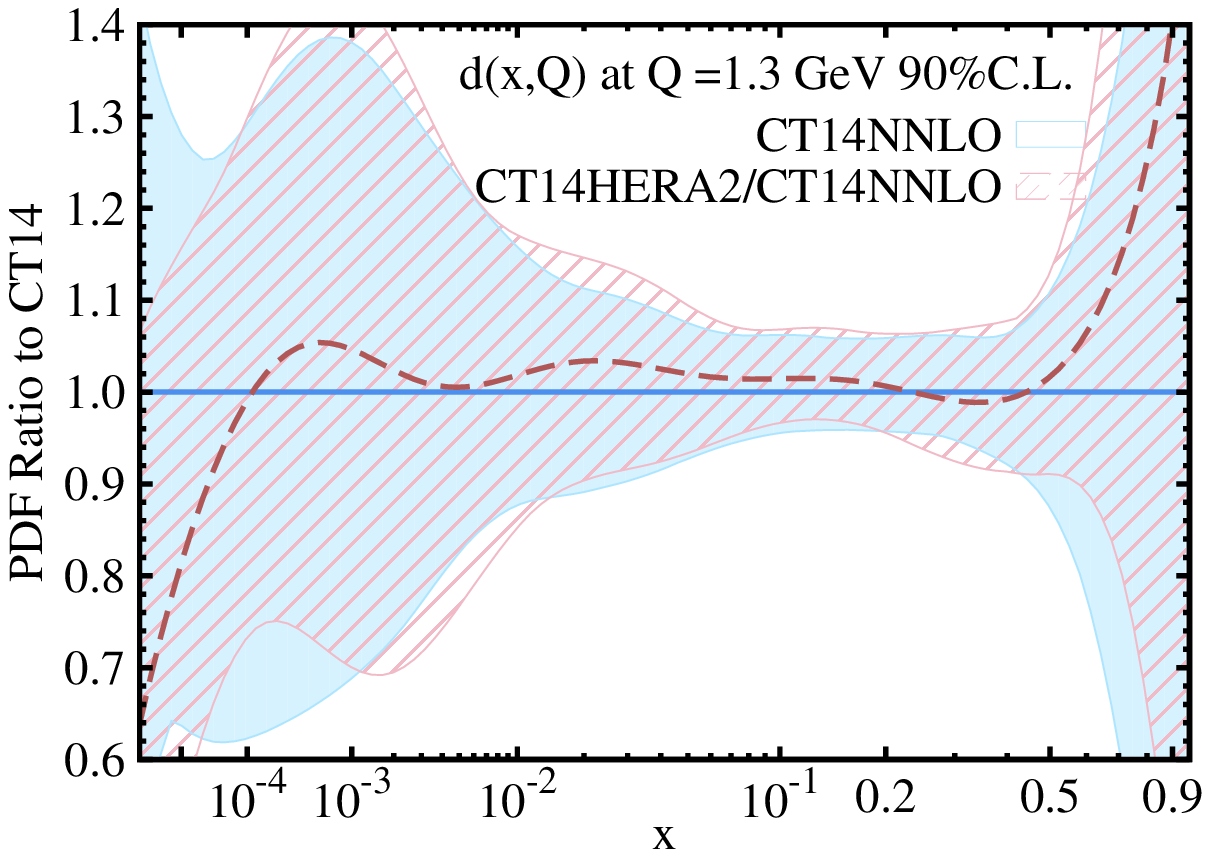}
\includegraphics[width=0.47\textwidth]
{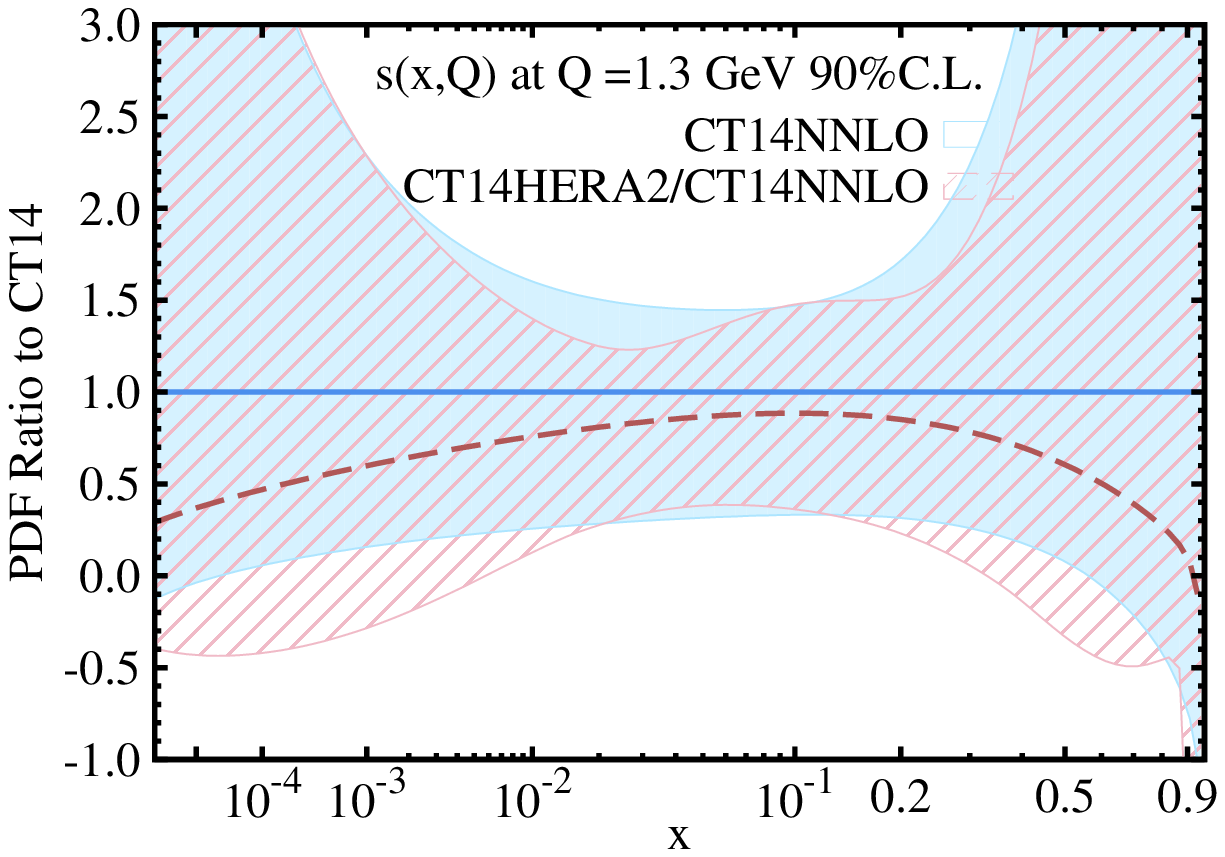}
\includegraphics[width=0.47\textwidth]
{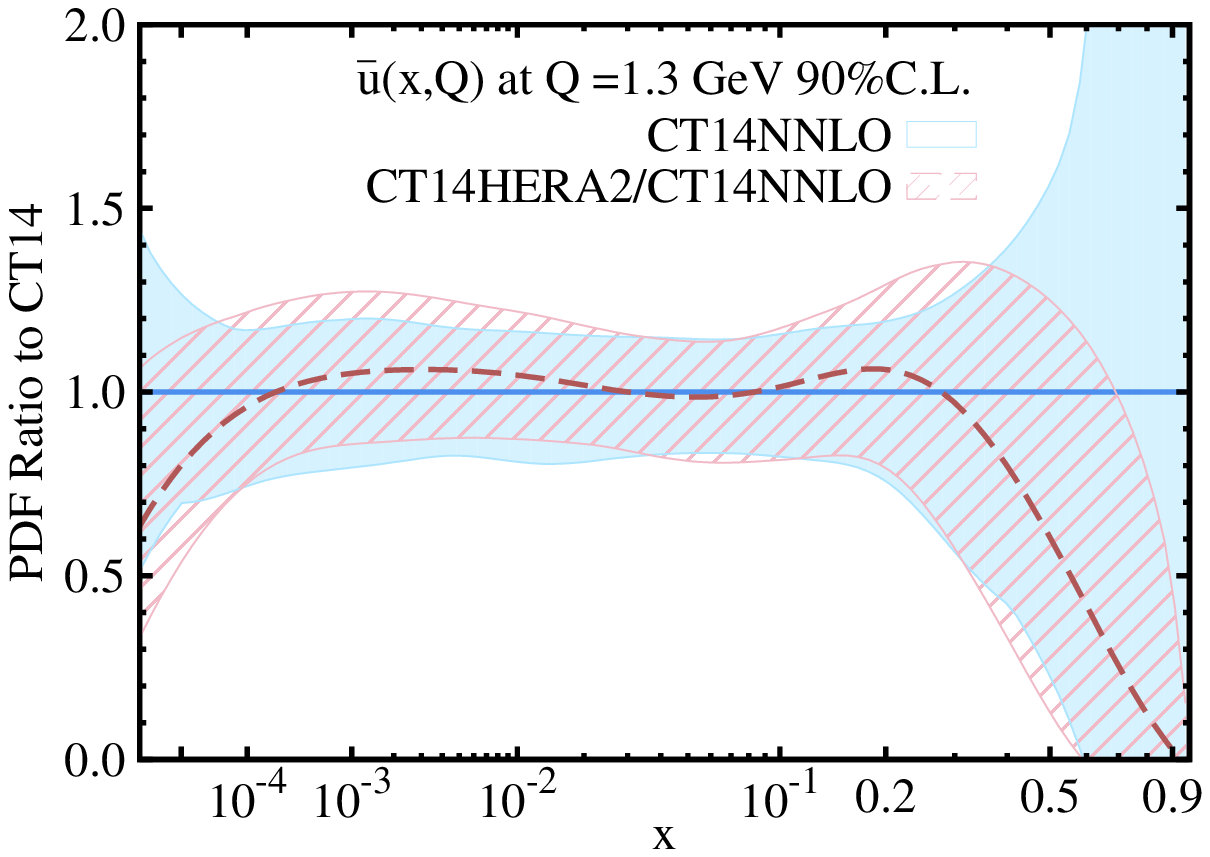}
\includegraphics[width=0.47\textwidth]
{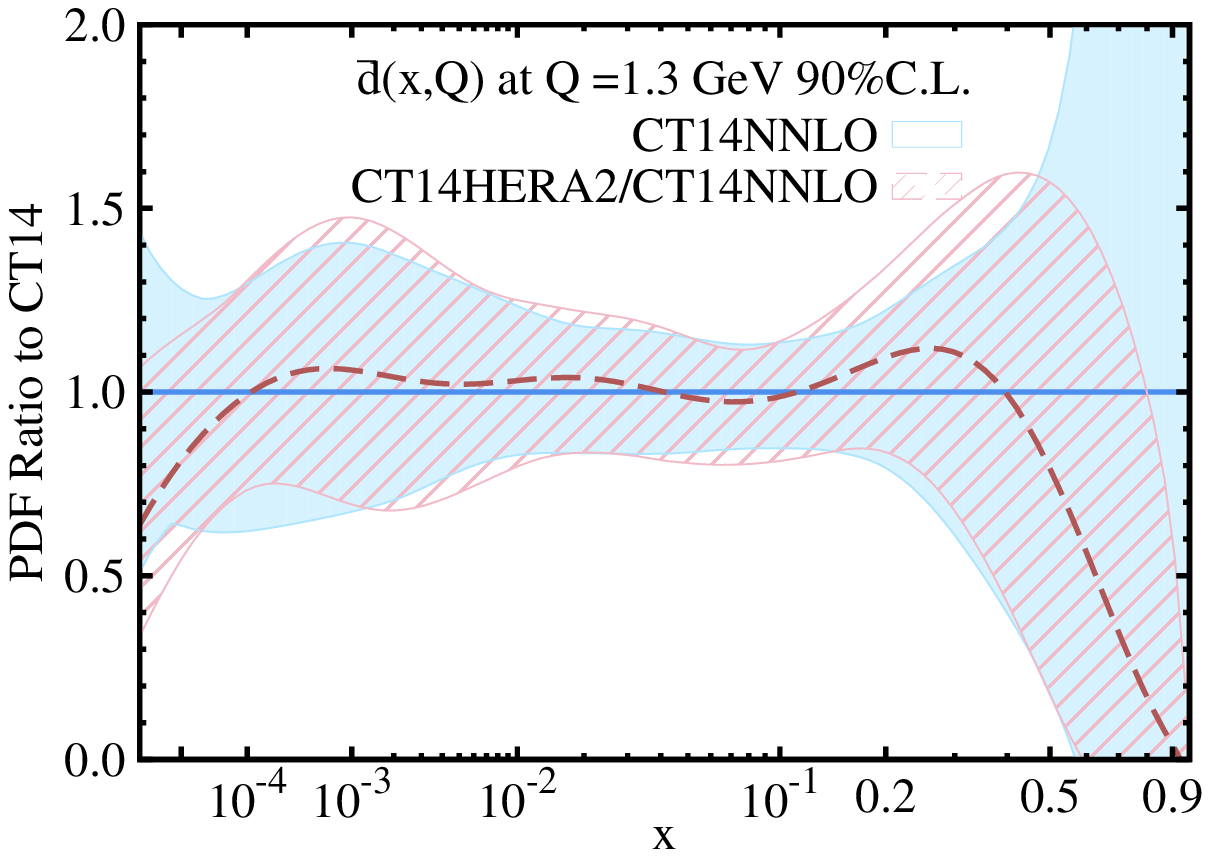}
\caption{
Comparison of CT14$_\textrm{HERA2}$ (dashed red) and CT14 (solid blue)
PDFs at $Q=1.3$ GeV.
Flavors $g, u, d, s, \bar{u},\bar{d}$ are shown.
The curves compare the central fits, plotted as ratios to CT14.
The uncertainty bands are
90\% C.L. uncertainties evaluated from the CT14 (shaded blue)
and CT14$_\textrm{HERA2}$ (hatched red) error ensembles;
both error bands are normalized to the corresponding central CT14 PDFs.
All PDFs are from the NNLO QCD analysis.
\label{fig:ct14h2PDFs}}
\end{figure}

\begin{figure}[htp]
\includegraphics[width=0.47\textwidth]
{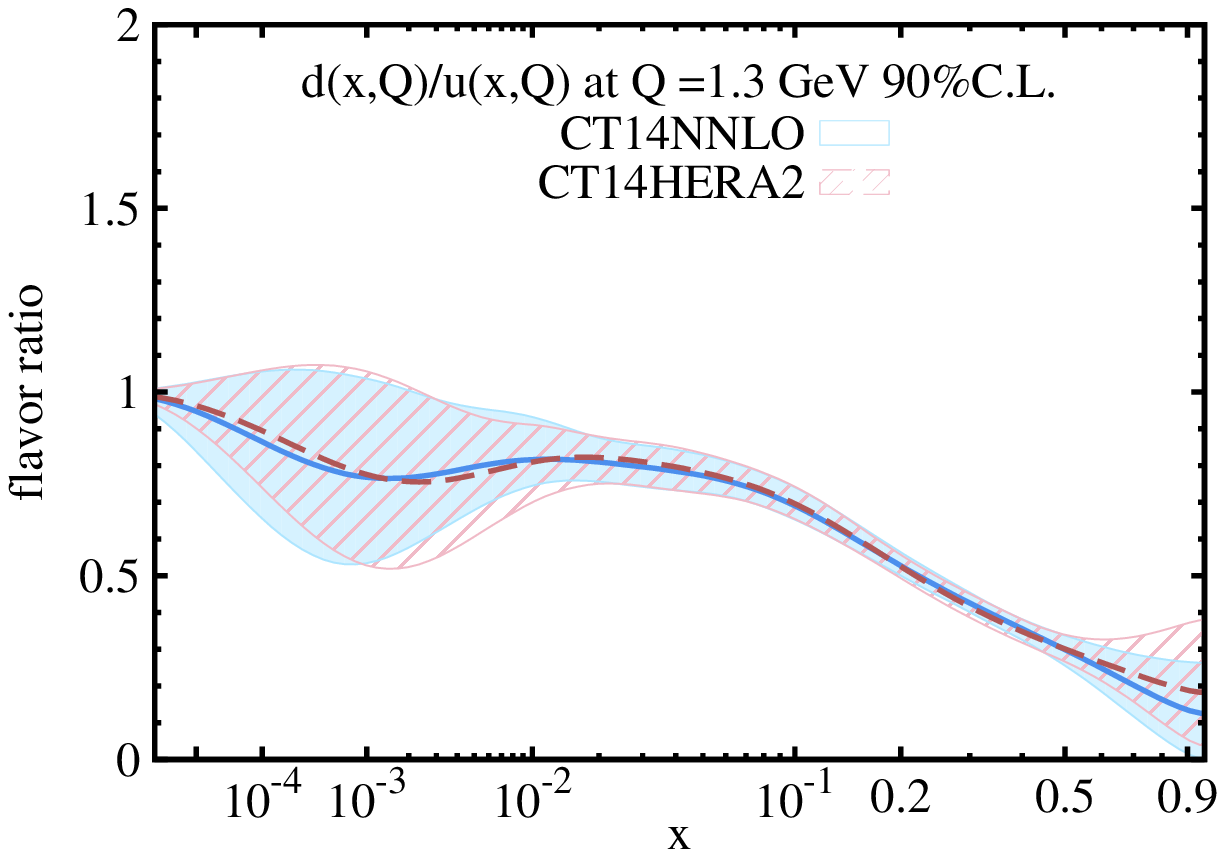}
\includegraphics[width=0.47\textwidth]
{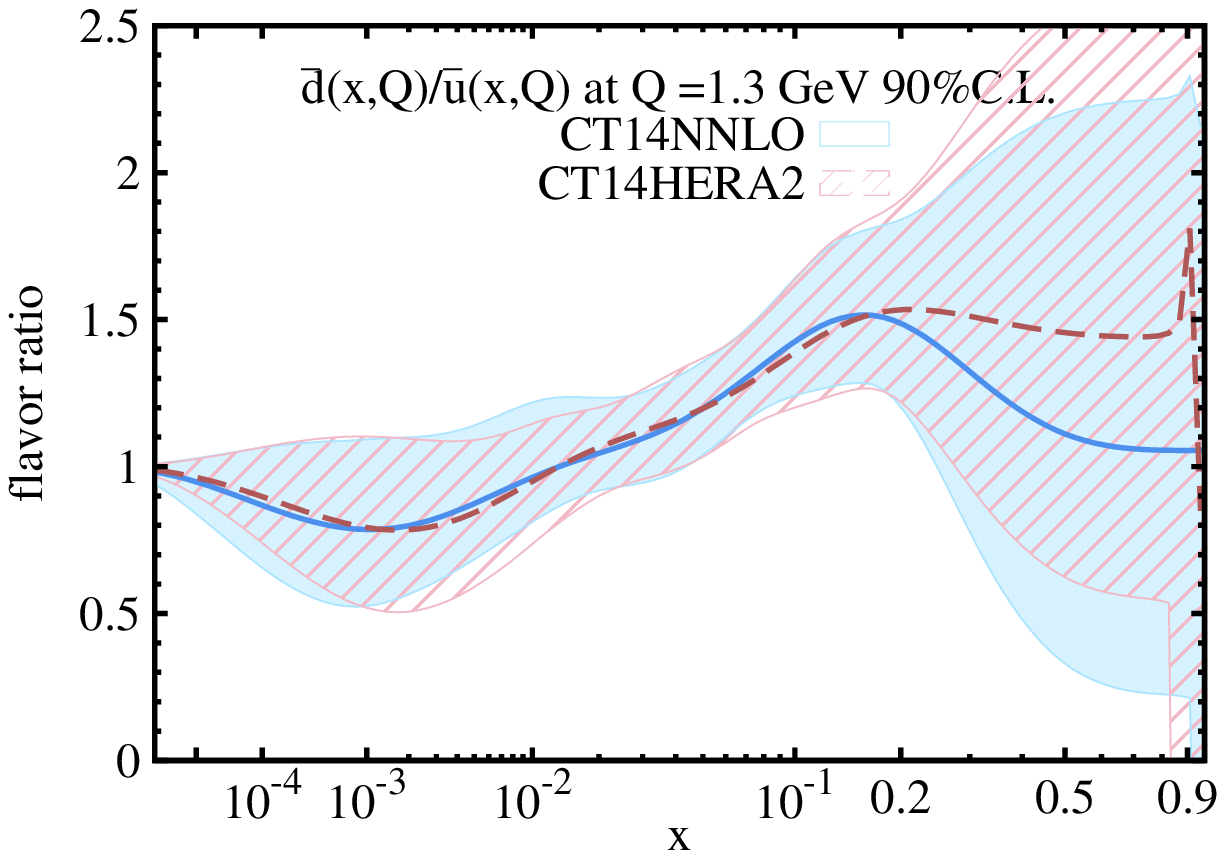}
\begin{center}
\includegraphics[width=0.47\textwidth]
{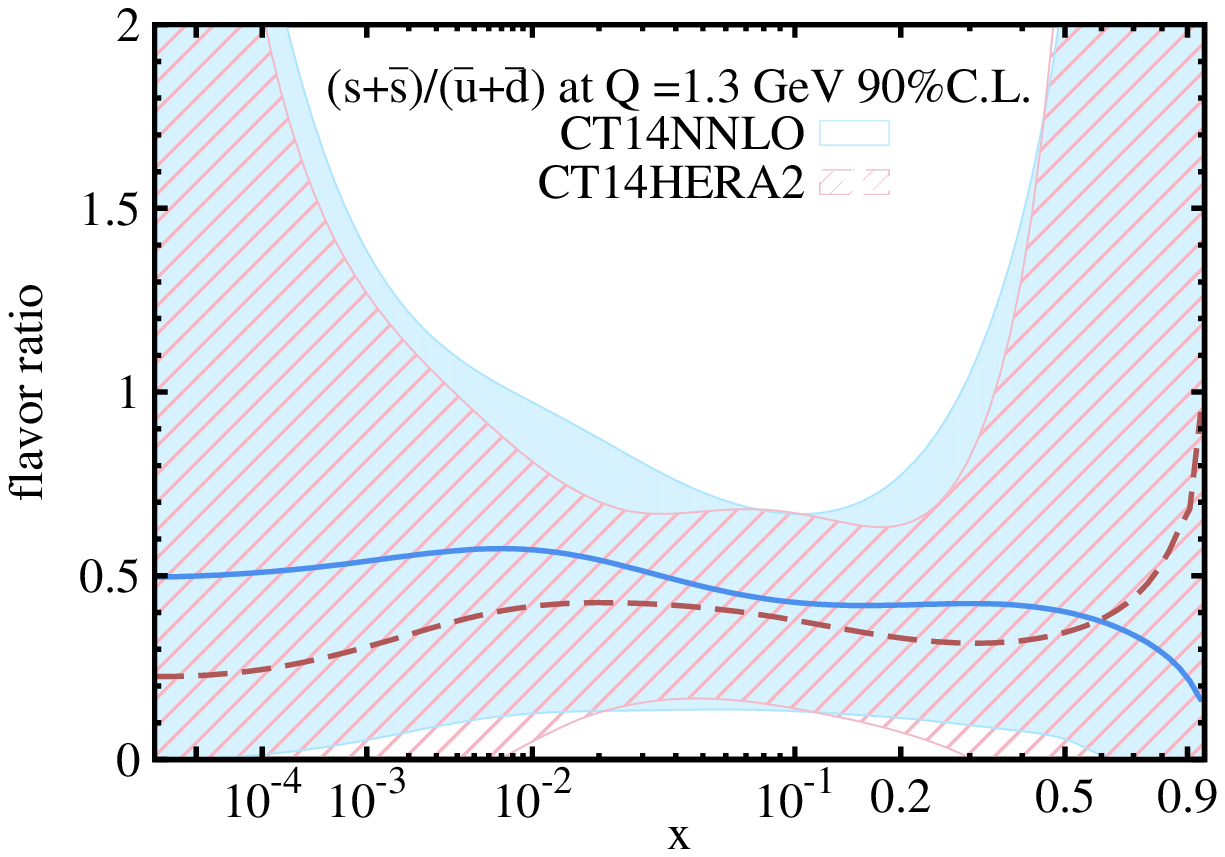}
\end{center}
\caption{Comparison of 90\% C.L. uncertainties on the ratios
$d/u$, $\bar{d}/\bar{u}$ and $(s+\bar{s})/(\bar{u}+\bar{d})$
at $Q = 1.3$ GeV.
The error bands are for the CT14 (solid blue)
and CT14$_\textrm{HERA2}$ (dashed red) error ensembles.
All PDFs are from the NNLO QCD analysis.
\label{fig:ct14h2RATs}}
\end{figure}

Figures~\ref{fig:ct14h2PDFs} and \ref{fig:ct14h2RATs} show plots
where CT14$_\textrm{HERA2}$ (dashed red) is compared to CT14 (solid blue),
including error bands. Some comments about this comparison are
listed below.

 \begin{itemize}

\item{The central value of the
CT14$_\textrm{HERA2}$ gluon in the range $10^{-2} \lesssim x \lesssim 0.2$
is almost unchanged compared to CT14;
it is larger by about 30\% at $x \approx 10^{-4}$,
by a larger factor for $x > 0.5$, and it
is smaller by about 10\% at $x \approx 0.3$.
}

\item{The up and down quarks are
generally slightly larger than (but close to) CT14 in the range $10^{-2}\lesssim x \lesssim 0.5$,
where the CT14$_\textrm{HERA2}$ uncertainty band is comparable to that of CT14;
whereas they are both systematically larger by about 5\%
in the intermediate region of $10^{-4}\lesssim x \lesssim 10^{-2}$.
The CT14$_\textrm{HERA2}$/CT14 ratio decreases at $x \lesssim 10^{-4}$
in both cases.
The down quark increases at $x>0.5$, while the up quark decreases slightly
at $x \approx 0.5$.
The slow oscillations in $d(x,Q_{0})$ reflect the behavior of
Bernstein polynomials in Eq. (1).
}

\item{The strange quark central PDF is reduced over the entire
  $x$ range, mainly due to the change of freeing one shape parameter for describing the strange (anti)quark PDF;
  but this reduction is statistically insignificant and
  completely within the uncertainty of the previous PDF ensemble.
In particular a reduction of approximately $-50$\% is observed at both
$x\lesssim 10^{-3}$ and $x\gtrsim 0.5$.
}

\item{The changes in $\bar{u}$ and $\bar{d}$ quarks share similar features.
These PDFs are almost unchanged for $10^{-2} \lesssim x \lesssim 0.2$.
The $\bar{u}$ quark PDF increases by about 10\% at $x$ around 0.2,
and the $\bar{d}$ quark PDF similarly increases at $x$ around 0.3.
Both the $\bar{u}$ and $\bar{d}$ quarks, similar to
the $s$ quark, decrease by large factors for $x \gtrsim 0.4$, where
the gluon and down quark PDFs increase, as a consequence of
the momentum sum rule.
It is important to keep in mind that at $x > 0.5$ the antiquark PDFs
take very small values, their
behavior is very uncertain and strongly depends on the parametrization
form.
}

\item {The individual PDF uncertainties do not change appreciably,
except in the unconstrained $x$ regions.
}

\item{We have verified that the change
seen in gluon, up and down quark PDFs mainly arises from
replacing the HERA1 data (in CT14 analysis) by the HERA2 data
(in CT14$_\textrm{HERA2}$ analysis). This was explicitly checked
by comparing CT14 PDFs to
the result of yet another new fit in which we used the exact same
setup as that in the
CT14 global analysis, but with the HERA1 data replaced by the
HERA2 data.
}

\end{itemize}

Now we turn to certain {\em ratios} of PDFs.
Figure~\ref{fig:ct14h2RATs} shows the most relevant
effects of the HERA2 data on the PDF ratios
at $Q_0=1.3$ GeV.
Comparing CT14$_\textrm{HERA2}$ to CT14 we observe the following.

\begin{itemize}

\item{The ratio $d/u$ remains approximately the same for
CT14$_\textrm{HERA2}$ and CT14,
in both the central value and uncertainty, for all values of $x$.}

\item{The ratio $\bar{d}/\bar{u}$ at $x\lesssim 0.1$ is about the same
for CT14$_\textrm{HERA2}$ and CT14, with compatible uncertainties.
However, it is larger for CT14$_\textrm{HERA2}$ as $x$ increases
beyond 0.2, despite having a large uncertainty.
We note that this change mainly arises from using the more
flexible parametrization in the strange quark PDF.}
An interesting feature is that $\overline{d}/\overline{u}$
is greater than 1 for \ZcthZ at large $x$ region.

\item{The strange quark fraction
$R_{s} = (s+\bar{s})/(\bar{u}+\bar{d})$
is an important PDF ratio that has been discussed recently
in several QCD analyses~\cite{Aad:2012sb,Samoylov:2013xoa,Chatrchyan:2013uja,Alekhin:2014sya, Dulat:2015mca}.
As done in the CT14 global analysis, we assume that $s$ and $\bar{s}$ PDFs
are the same at the initial $Q_0$ scale.
We find that the value of $R_{s}$ for CT14$_\textrm{HERA2}$
is smaller than for CT14 in the $x$
range from $10^{-4}$ to $0.5$.
This is mainly because the strange quark PDF decreases
when going from CT14 to \ZcthZ, as discussed above.
}
\end{itemize}

\section{Discussion and Conclusions}
\label{sec:conclusions}

In this paper, we have presented the \ZcthZ parton
distribution functions,
constructed from a global analysis of QCD
that uses the HERA Run I and II combined data set on $e^\pm p$ deeply
inelastic scattering~\cite{Abramowicz:2015mha}.
This compendium of 20 years of HERA data,
reconciled as well as possible,
including comparative analysis of systematic errors
from the two collaborations, H1 and ZEUS,
provides the most comprehensive information about DIS available today.
A comparison of the current QCD analysis of this data (HERA2) to the
CT14 global analysis of the previous generation of HERA data (HERA1)
yields important insights about the structure of the nucleon, at the
highest precision achieved.

The main purpose of the paper is to examine the quality of agreement
of perturbative QCD predictions with the HERA2 data and discuss
the impact of these data on the PDFs and their uncertainties used for
a variety of LHC applications. We conclude that the \ZcthZ and CT14 PDFs,
do have some differences.
However, the differences are smaller than the PDF uncertainties
of the standard CT14 analysis.

Some specific features of the \ZcthZ PDFs are elucidated
in the paper.

\begin{itemize}
\item{Figure 2 shows values of $\chi^{2}/N_{pts}$ for the HERA2 data.
  $\chi^2/N_{pts}$ is marginally smaller in the NLO analysis than
  at NNLO, but the difference is clearly negligible.
In either case, $\chi^{2}$ decreases as HERA2 data is
included with increasing weight, at about the same rate
for NLO and NNLO.}

\item{Figures 4 and 5 show that HERA2 data slightly modify the
$g$, $d$, and $u$ PDFs. The $s$ PDF decreases,
mainly due to the
use of a slightly more flexible parametrization for the
strange quark PDF.
The $\bar u$ and $\bar d$ PDFs decrease at large $x$,
where $g$ and $d$ PDFs increase, so as to satisfy
the momentum sum rule.
The most significant effects of the HERA2 data
in the \ZcthZ analysis are seen in
the ratio of $\overline{d}/\overline{u}$ which is greater than 1
for very large $x$, although this change is much less than
the size of the error band.
Also, the strangeness fraction $R_{s}$ is
roughly 20\% smaller than the standard CT14 $R_{s}$
for the intermediate range of $x$. This is mainly caused
by the reduction in the strange quark PDF.
}

\end{itemize}

Because the CT14 and \ZcthZ PDFs agree well within the PDF errors,
we do not expect noticeable
differences in their predictions for
experimental
observables at the LHC. We have explicitly checked that
using \ZcthZ and CT14 PDFs at NNLO gives almost the same predictions for
the cross section for $W^\pm$ and $Z$
production~\cite{Aad:2016naf,CMS:2015ois,Aad:2011dm, CMS:2011aa, Chatrchyan:2014mua},
as well as the
associated $W^\pm$ and charm production~\cite{Chatrchyan:2013uja},
at the LHC energies.

In future CT analyses we may employ the HERA2 combined data
as an important part of the global data set, together with the
new LHC data that will be published,
such as low- and high-mass Drell-Yan processes
and top quark differential distributions.
For the present, we continue to recommend
CT14 PDFs for the analysis of LHC Run 2 experiments.
However, we make the
CT14$_\textrm{HERA2}$ PDFs available
in the LHAPDF format for specialized studies, such as those that are
sensitive to behavior of strange (anti)quark PDFs.

\begin{acknowledgments}
This research was supported in part by the National Science Foundation
under Grants No. PHY-1410972 and No. PHY-1417326; by the
U.S. Department of Energy under Award No. DE-AC02-06CH11357 and Grants
No. DE-SC0013681 and No. DE-SC0010129; by the
 National Natural Science Foundation of China under Grant
 No. 11465018;
and by the Lancaster-Manchester-Sheffield Consortium
for Fundamental Physics under STFC Grant No.~ST/L000520/1.

\end{acknowledgments}

\bibliographystyle{h-elsevier3}

\end{document}